\documentclass[12pt,preprint]{aastex}
\newcommand\msol{{\cal M_{\odot}}}

\newcommand\teff{{T_{\rm eff}}}
\newcommand\amlt{{\alpha_{\rm MLT}}}

\newcommand\lta{\mathrel{\hbox{\raise 0.6 ex \hbox{$<$}\kern
                   -1.8 ex\lower .5 ex\hbox{$\sim$}}}}
\newcommand\gta{\mathrel{\hbox{\raise 0.6 ex \hbox{$>$}\kern
                   -1.7 ex\lower .5 ex\hbox{$\sim$}}}}
\newcommand\minplu{\mathrel{\hbox{\raise 0.7 ex \hbox{$-$}\kern
                   -1.7 ex\lower .1 ex\hbox{$+$}}}}
\newcommand{\Hipp}{{\it Hipparcos}}
\newcommand{\HST}{{\it HST}}
\newsavebox\uscorebox
\shortauthors{VandenBerg et al.}
\shorttitle{Ages of Three Field Halo Subgiants}
 
\begin{document}
 
\title{THREE ANCIENT HALO SUBGIANTS: PRECISE PARALLAXES, COMPOSITIONS, AGES,
AND IMPLICATIONS FOR GLOBULAR CLUSTERS\altaffilmark{1,2,3}}

\author{Don A.~VandenBerg}
\affil{Department of Physics \& Astronomy, University of Victoria,
       P.O.~Box 1700 STN CSC, Victoria, B.C., V8W~2Y2, Canada}
\email{vandenbe@uvic.ca}

\author{Howard E.~Bond}
\affil{Department of Astronomy \& Astrophysics, Pennsylvania State University,
       University Park, PA\ 16802, USA; Space Telescope Science Institute,
       3700 San Martin Drive, Baltimore, MD\ 21218, USA}
\email{heb11@psu.edu}

\author{Edmund P. Nelan}
\affil{Space Telescope Science Institute, 3700 San Martin Drive, 
       Baltimore, MD\ 21218, USA}
\email{nelan@stsci.edu}

\author{P.~E.~Nissen}
\affil{Stellar Astrophysics Centre, Department of Physics \& Astronomy, 
       Aarhus University, Ny Munkegade 120, DK-8000 Aarhus C, Denmark}
\email{pen@phys.au.dk}

\author{Gail H.~Schaefer}
\affil{The CHARA Array of Georgia State University, Mount Wilson Observatory,
       Mount Wilson, CA\ 91023, USA}
\email{schaefer@chara-array.org}

\author{Dianne Harmer}
\affil{National Optical Astronomy Observatories, 950 North Cherry Avenue,
       Tucson, AZ\ 85726, USA}
\email{diharmer@noao.edu}

\altaffiltext{1}{Based in part on observations made with the NASA/ESA {\it
Hubble Space Telescope}, obtained by the Space Telescope Science Institute.
STScI is operated by the Association of Universities for Research in
Astronomy, Inc., under NASA contract NAS5-26555.}
\altaffiltext{2}{Based in part on observations obtained at the Kitt Peak
National Observatory and Cerro Tololo Interamerican Observatory, National
Optical Astronomy Observatory, which is operated by the Association of
Universities for Research in Astronomy, Inc., under cooperative agreement
with the National Science Foundation.}
\altaffiltext{3}{Based in part on observations collected at the La Silla
Paranal Observatory, ESO, Chile.}

\begin{abstract}
The most accurate ages for the oldest stars are those obtained for nearby halo
subgiants, because they depend almost entirely on just the measured parallaxes
and absolute oxygen abundances.  In this study, we have used the Fine Guidance
Sensors on the {\it Hubble Space Telescope} to determine trigonometric
parallaxes, with precisions of 2.1\% or better, for the Population II
subgiants HD\,84937, HD\,132475, and HD\,140283.  High quality spectra have
been used to derive their surface abundances of O, Fe, Mg, Si, and Ca, which
are assumed to be 0.1--0.15 dex less than their initial abundances due to the
effects of diffusion.  Comparisons of isochrones with the three subgiants on
the $(\log\teff,\,M_V)$-diagram yielded ages of $12.08 \pm 0.14$, $12.56 \pm
0.46$, and $14.27 \pm 0.38$ Gyr for HD\,84937, HD\,132475, and HD\,140283, in
turn, where each error bar includes only the parallax uncertainty.  The total
uncertainty is estimated to be $\sim \pm 0.8$ Gyr (larger in the case of the
near-turnoff star HD\,84937).  Although the age of HD\,140283 is greater than
the age of the universe as inferred from the cosmic microwave background by
$\sim 0.4$--0.5 Gyr, this discrepancy is at a level of $< 1\,\sigma$.
Nevertheless, the first Population~II stars apparently formed very soon after
the Big Bang.  (Stellar models that neglect diffusive processes seem to be
ruled out as they would predict that HD\,140283 is $\sim 1.5$ Gyr older than
the universe.)  The field halo subgiants appear to be older than globular
clusters of similar metallicities: if distances close to those implied by the
RR Lyrae standard candle are assumed, M\,92 and M\,5 are younger than
HD\,140283 and HD\,132475 by $\sim 1.5$ and $\sim 1.0$ Gyr, respectively. 
\end{abstract}
 

\section{Introduction}
\label{sec:intro}

Stars with extremely low metal abundances are of particular astrophysical and
cosmological interest because they probe very early times in the evolution of
the universe and its galactic components.  However, it is very difficult to
determine their absolute ages to within $\sim 15$--20\%.  Nearly all of the
metal-poor stars that can be age-dated reliably are found in globular clusters
(GCs) or dwarf galaxies, but even when tight, well-defined color-magnitude
diagrams (CMDs) are available for them, the ages that are inferred from fits of
isochrones to the observed CMDs depend critically on the adopted distances and
chemical abundances.  For instance, a $\pm 0.10$ mag uncertainty in the assumed
distance modulus translates to a change in the estimated age by $\sim\,\minplu
1$ Gyr --- which is also implied by the assumption of an oxygen abundance that
is altered by $\delta(\log\,N) \approx \pm 0.3$ (see, e.g., \citealt{vbd12}).
In view of such sensitivities, it is not surprising that recent estimates of
the ages of well-studied GCs like M\,92 (e.g., \citealt{dbb10},
\citealt{vbl13}) and 47 Tucanae (e.g., \citealt{dsa10}, \citealt{hka13}) range
over as much as 2--3 Gyr.

Ages of considerably higher accuracy are possible for field
Population II stars {\it if} they are close enough that accurate and precise
trigonometric parallaxes can be obtained for them {\it and} they are in the
subgiant phase of evolution.  Isochrones for different ages are most widely
separated in the region of a CMD between the turnoff (TO) and the base of the 
red-giant branch (RGB); i.e., the luminosities of stars, at a fixed effective
temperature ($\teff$) or color, are predicted to depend most strongly on age
when they are evolving along what is usually referred to as the subgiant branch
(SGB).  As a consequence, the ages of such stars can be determined directly
from their locations relative to isochrones that have been computed for their
observed chemical abundances.  Although the ramifications of temperature and/or
color uncertainties are not insignificant, such comparisons between theory and
observations benefit from the fact that the slope of the SGB is relatively
shallow on the $(\log\teff,\,\log\,L)$-diagram and most CMDs.  A further big
advantage of studying nearby SGB stars, as opposed to their counterparts in
considerably more distant systems, is that much higher resolution, higher S/N
spectra may be used to determine their metal abundances.  Moreover, the
foreground reddenings are small and they can be determined precisely from the
interstellar Na\,{\sc i}\,D lines.

HD\,140283, which has long been known to have [Fe/H] $\sim -2.4$ (e.g.,
\citealt{nca04}, \citealt{b07}, \citealt{crm10}), is the most famous
example\footnote{For a discussion of the role of HD\,140283 in the history of
astronomy, see \citet{bnv13}.} of a solar-neighborhood subgiant that provides
a direct constraint on the ages of
field stars of similar metallicity due to its auspicious location on the H-R
diagram.  The parallax that was initially derived for it, $\pi = 17.44 \pm 0.97$
mas, from measurements taken with the {\it Hipparcos} astrometric satellite
(\citealt{plk97}) yielded $M_V \approx 3.30 \pm 0.12$, depending on the adopted
apparent magnitude, reddening, and the statistical corrections that were applied
(e.g., \citealt{pmt98}, \citealt{cgc00}).  The corresponding age of HD\,140283
was determined to be $13.5 \pm 1.4$ Gyr (where the error bar takes into account
only the distance uncertainty) by \citet{vrm02}, using isochrones based on
University of Montreal stellar models that treated all of the important
diffusive processes (\citealt{rmr02}).  At nearly the same time, the age of the
universe was found to be $13.7 \pm 0.2$ Gyr from the analysis by \citet{svp03}
of first-year {\it Wilkinson Microwave Anisotropy Probe (WMAP)} observations of
the cosmic microwave background (CMB).  The similarity of these results suggests
that HD\,140283 has the potential to provide an interesting test of stellar
physics and cosmology, and to contribute to our understanding of the first
epochs of star formation in the early universe, {\it if} its age can be pinned
down much more precisely.

To this end, we began a program in 2003 August to improve upon the {\it
Hipparcos} parallaxes of HD\,140283 and the two other nearest metal-deficient
subgiants, HD\,84937 and HD\,132475, using
the Fine Guidance Sensors (FGSs) on the {\it Hubble Space Telescope (HST)}.
[HD\,140283 and HD\,132475, which differ in [Fe/H] by $\sim 1$ dex, are arguably
the best candidates for such a study because they are located approximately
midway between the TO and the lower RGB; see \citet[his Figs.~9 and
32]{van00}.  HD\,84937 lies closer to the TO, and in fact, it has sometimes been
considered to be a blue straggler; e.g., \citet{bm71}.]  As reported by
\citet{bnv13}, our reduction of the observations
of HD\,140283 that were obtained between 2003 and 2011 resulted in a value of
$\pi = 17.15 \pm 0.14$ mas, which agrees remarkably well with the value derived
by \citet[$\pi = 17.16 \pm 0.68$ mas]{vl07} in his re-analysis of the original
{\it Hipparcos} data, but with a factor of 5 improvement in the precision of
the result.  On the assumption of $V = 7.205 \pm 0.02$ (\citealt{crm10}) and
$E(B-V) = 0.000 \pm 0.002$ (\citealt{mcr10}), the FGS parallax yielded an
absolute magnitude $M_V = 3.377 \pm 0.027$.  According to the many published
spectroscopic studies that were consulted (see the Bond et  al.~paper for
references), HD\,140283 has [Fe/H] $= -2.4 \pm 0.10$ and [O/H] $= -1.67 \pm
0.15$: these values were adjusted by $+0.1$ and $+0.13$ dex, respectively, to
compensate for the predicted effects of diffusion and turbulent mixing on its
surface abundances over its lifetime (\citealt{rmr02}).  From up-to-date
isochrones for the resultant {\it initial} metal abundances, the age of
HD\,140283 was found to be $14.46 \pm 0.8$ Gyr, where the [O/H] and $\teff$
uncertainties contribute more to the total error bar than the distance
uncertainty (which amounts to $\pm 0.31$ Gyr).

The current best estimate of the age of the universe is $13.82 \pm 0.06$ Gyr,
based on observations of the CMB using the {\it Planck} satellite
(\citealt{aaa13}), in excellent agreement with the latest {\it WMAP} derivation
of $13.77 \pm 0.06$ Gyr (\citealt{blw13}).  Recent simulations (e.g.,
\citealt{rgs12}; \citealt{smb14}) suggest that the oldest Population II stars
probably formed $\sim 0.2$--0.3 Gyr after the Big Bang, depending on how
quickly the gas from the first (Population III) supernovae was able to cool
and condense, as well as such factors as the severity of the Pop.~III feedback.
Hence, the age that Bond et al.~(2013) obtained for HD\,140283 is about 0.8 Gyr
older than its expected maximum age.  Although 13.5--13.6 Gyr is only just
outside the $1\,\sigma$ error bar on the age of HD\,140283 (when all sources of
error are considered), which is not a severe inconsistency, it is clearly
worthwhile to take a second, hard look at all of the factors that play a role
in the age determination, including the parallax, the adopted metal abundances,
and the stellar models.  In particular, we decided to carry out our own analysis
of high-resolution spectra of HD\,140283 (and the other subgiants in our sample)
rather than rely on published (sometimes discordant) chemical abundance
determinations.

The next section provides an overview of the procedures used to derive
trigonometric parallaxes using the FGSs and the results that we obtained for
our target stars.  Our spectroscopic analyses, the stellar models that have
been used, and the determination of the ages of the subgiants are fully
described in sections \ref{sec:atm}, \ref{sec:models}, and \ref{sec:ages},
respectively.  The implications of our findings for the distances, ages, and
[Fe/H] scales of globular clusters is the subject of \S~\ref{sec:gcs}, which
also includes some discussion concerning the RR Lyrae $M_V$ versus [Fe/H]
relationship.  The main conclusions of this study are summarized in
\S~\ref{sec:sum}.

\section{Hubble Space Telescope FGS Astrometry}
\label{sec:pi}

\subsection{Observations and Data Analysis}
\label{subsec:fgs}

Our astrometric measurements were made with the FGSs onboard the {\it HST}.
They are a set of three interferometers that can sequentially measure precise
positions of a target star and several surrounding astrometric reference stars
with one FGS (FGS1r), while the other two (FGS2 and FGS3) are locked on guide
stars in order to maintain telescope pointing.  Our FGS observations of
each halo subgiant consisted of two \HST\/ orbits each at five epochs, taken
near the biannual dates of maximum parallax factor, for a total of
10 \HST\/ orbits per target (11 for HD\,140283). The observations were made
as follows: HD\,84937 between 2003 November and 2005 November; HD\,132475
between 2003 August and 2006 February; and HD\,140283 between 2003 August and
2006 March, with an additional single orbit of observation in 2011 March.

During each \HST\/ orbit, we used FGS1r in ``Position'' mode to measure the
relative positions of the subgiant and six or seven faint neighboring background
reference stars. Each reference star was observed for about 30~s, three to four
times per orbit, interspersed with five or more measurements of the subgiant.
The data gathered over the course of an orbit were processed and calibrated as a
single ``plate." (The repeated exposures of the same stars are used to model and
remove telescope pointing drift resulting from the changing thermal environment
as the \HST\/ orbits the Earth.)  Guide-star telemetry from the two guiding FGSs
was used to remove high-frequency spacecraft pointing jitter from the FGS1r
astrometric data. Because of refractive components in the FGS optical chain,
color-dependent corrections to the astrometry were applied as functions of $B-V$
color for each star (see \citealt[their \S~3.3 for details]{nb13}). Finally, the
FGS data were corrected for differential velocity aberration and geometric
distortion (``optical field angle distortion'') to produce the astrometric
measurements from each \HST\/ orbit.

Data from the ten or eleven orbits were combined using an updated version of the
least-squares overlapping-plate GAUSSFIT program (\citealt{jfa88}),
which solves for the plate constants (rotation, translation, and scale) along
with the parallax and proper motion (PM) of each star. However, since the FGS
measures only relative positions, the PMs and estimated parallaxes of the
reference stars must be input as priors to the model in order to obtain an
absolute parallax and PM of the subgiant. We estimated the input reference-star
parallaxes based on ground-based photometry and spectroscopic observations (see
next subsection). Their input PMs were obtained from the PPMXL (\citealt{rds10})
or UCAC4 (\citealt{zfg13}) catalogs; we chose the input catalog for each
field that yielded the lower $\chi^2$ value, which proved to be PPMXL for
the HD\,84937 field and UCAC4 for the other two.  These reference-star
data are provided to the model as observations with errors, not as fixed
values. This allows GAUSSFIT freedom to adjust the values within
their stated errors to find the overall best $\chi^2$ solution.  We assumed a
20\% uncertainty in the distance of each reference star, corresponding to a
$\pm 0.4$~mag uncertainty in their absolute magnitudes.  (An overview of
parallax measurements with the FGSs can be found in \citealt{nm02}.)

% We inspected the results for outlying individual observations and for very
% discrepant reference stars, for example those whose spectroscopic parallax
% differs significantly from the \HST\/ result reported by GAUSSFIT\null.  If
% warranted, such data are removed from the analysis and the model is rerun to
% convergence.

The input ground-based PMs generally have large uncertainties, relative to the
high precision of the \HST\/ measurements. This results in elevated residuals in
all of the other quantities computed by the model. To mitigate this, we first
executed the model with the ground-based measurements and errors, and then used
the resulting FGS PMs and errors as input to the next iteration. We found that,
while this does not appreciably change the PMs after the first iteration, the
resulting $\chi^2$ and residuals are generally reduced by a few tens of percent.

The $V=7.2$ brightness of HD\,140283 necessitated the use of the F5ND
attenuator (required for targets brighter than $V=8$) on the subgiant, while
the fainter reference stars were observed using the F583W spectral element. 
This introduces a cross-filter wedge effect that shifts the
position of HD\,140283 relative to the field stars in a manner similar to
parallax. Fortunately, this cross-filter shift (on the order of 8~mas) is well
calibrated near the two locations in the FGS1r field of view where HD\,140283
was observed over the course of the program. Nonetheless, there remains a
$\sim 0.2$~mas systematic uncertainty in the cross-filter calibration, which
must be added in quadrature to the statistical errors computed by GAUSSFIT\null.
This had not been included in the uncertainties reported for HD\,140283 by 
\citet{bnv13}, but is incorporated now. For the other two targets,
HD\,84937 and HD\,132475, the F5ND filter was not used for any of the
observations.

\subsection{Reference-Star Distance Estimation}
\label{subsec:ref}

We made distance estimates for the reference stars based on ground-based
spectroscopy and photometry. For spectral classification, digital spectra in all
three fields were obtained by us with the WIYN 3.5-m telescope and Hydra
spectrograph at Kitt Peak National Observatory (KPNO) in 2004 and 2007, and by
service observers on the 1.5-m SMARTS telescope\footnote{SMARTS (see
{\tt http://www.astro.yale.edu/smarts}) is the Small \& Moderate Aperture
Research Telescope System.} and Ritchey-Chretien
spectrograph at Cerro Tololo Interamerican Observatory (CTIO) in 2003 through
2005. The classifications were then determined through comparisons with a
network of MK standards obtained with the same telescopes, assisted by
equivalent-width measurements of lines sensitive to temperature and luminosity.

Photometry of the reference stars in the Johnson-Kron-Cousins {\it BVI\/} system
was obtained with the SMARTS 1.3-m telescope at CTIO, using the ANDICAM CCD
camera, and calibrated to the standard-star network of \citet{lan92}. Each
field was observed by Chilean service observers on four different photometric
nights in 2003 through 2005.  Additional photometry was obtained by H.E.B.~with
the CTIO 0.9-m telescope on one night in 2001 and with the KPNO 0.9-m telescope
on two nights in 2007; and by D.H.~with the KPNO 2.1-m telescope on single
nights in 2003 and 2004.

Table~\ref{tab:tab1} gives the designations, coordinates, photometry,
and spectral types of the reference stars in columns 1 through~7. The
uncertainties in the magnitudes and colors, based on the internal agreement,
are typically about $\pm 0.004$ and $\pm 0.005$~mag, respectively. To
estimate the reddening of the reference stars (assumed to be the same for all
stars in each field, since their distances place them well beyond the dust
of the Galactic disk), we compared the observed $B-V$ color of each star with
the intrinsic $(B-V)_0$ color corresponding to its spectral type, and
calculated the average $E(B-V)$ for each field. The intrinsic colors were taken
from literature compilations\footnote{Available at
http://www.pas.rochester.edu/$^\sim$emamajek\slash
EEM\_dwarf\_UBVIJHK\_colors\_Teff.dat for dwarfs, and at URLs similar to
http://www.pas.rochester.edu\slash $^\sim$emamajek/spt/K0III.txt for subgiants
and giants. } assembled by E.~Mamajek. These averages agreed very well with
$E(B-V)$ values from the extinction maps of \citet[hereafter SF11]{sf11}, as
implemented at the NASA/IPAC
website\footnote{\url{http://ned.ipac.caltech.edu}}. Our adopted values of
$E(B-V)$ were 0.037, 0.100, and 0.130~mag for the background fields of
HD\,84937, HD\,132475, and HD\,140283, respectively. 

For the distance estimations, we used a purely empirical approach, which is
based on determining the luminosity class from the spectroscopic data, and then
using the dereddened photometry to estimate the absolute magnitude, $M_V$\null.
For subgiants and giants, we found the absolute magnitudes by interpolation in a
fiducial sequence [$M_V$ vs.\ $(V-I)_0$] for the old open cluster M\,67
(\citealt{san04}), which we took to be representative of the faint population at high
galactic latitudes ($b=  +45\fdg5, +31\fdg9$, and $+33\fdg6$, in turn,
for HD\,84937, HD\,132475, and HD\,140283).

For the reference stars classified as dwarfs, we followed the algorithm
described in \citet{bnv13}, which in brief is based on calibrations of the
visual absolute magnitude against $B-V$ and $V-I$ colors through polynomial fits
to a sample of 791 single main-sequence stars with accurate {\it BVI\/}
photometry and \Hipp\/ or USNO parallaxes of 40~mas or higher (provided online
by I.~N.~Reid\footnote{\url{http://www.stsci.edu/$\sim$inr/cmd.html}}).
This calibration includes a correction for metallicity, estimated from each
star's position in the $B-V$ vs.\ $V-I$ diagram.  We tested our algorithm by
applying it to 136 nearby stars with accurate parallaxes and a wide range of
metallicities listed by \citet{crm10}. We reproduced their known
absolute magnitudes with an {\it rms} scatter of only 0.28~mag. At the distances of
the reference stars, ranging mostly from about 500 to 1700~pc, this scatter
corresponds to parallax errors of 0.08 to 0.24~mas.

The final two columns in Table~\ref{tab:tab1} give the input estimated
parallaxes and associated errors for the reference stars, and the final
adjusted parallax that was output by the GAUSSFIT routine. In most cases,
the adjustments were quite small.

% HD 84937 	W BVI 040118 040119
% 		 040120 050207	
% 		E BVI 040118 040119
% 		 040120 050207      all 1.3m
% 
% 		uBVI 040412  2.1m by Di harmer
% 		uBVI 071027 poor kpno 0.9m by me
% 		uBVI 031206  2.1m by Di harmer
% 
% 
% HD 132475  	W BVI 030801 030802
% 		 030808	050317	
% 		E BVI 030801 030802
% 		 030808 050317
% 
% 		uBVI 040412  2.1m by Di harmer
% 
% HD 140283   	uBVI 010325 CTIO 0.9m by me
%                      070508 kpno 0.9m by me
% 
% 		W BVI 030801 030802
% 		 030808 050317	
% 		E BVI 030801 030802
% 		 030808 050317
% 
% 		uBVI 040412  2.1m by Di harmer

\subsection{Astrometric Results}
\label{subsec:ast}

Table~\ref{tab:tab2} presents our final astrometric results for the three halo
subgiants.  Columns~2 through 4 show the \Hipp\/ PM components and parallaxes,
and columns 5 through~7 show our FGS results. Note that we have reanalyzed the
FGS data for HD\,140283 that were presented by \citet{bnv13}, using the
identical procedures employed for the other two targets in the present paper.
Our results for this star are thus very slightly different from those published
in 2013. The error bar for HD\,140283 also now includes the uncertainty due
to use of the neutral-density filter, as discussed above in \S~\ref{subsec:fgs}.

The FGS PMs agree very well with those found by \Hipp\/. However, there are
slight offsets in the PM zero-points for the FGS measurements compared to the
absolute PMs of \Hipp, because the FGS reference frames are based on only about
half a dozen background stars chosen from the PPMXL or UCAC4 catalogs. These
catalogs have errors in the absolute PMs per star of several mas~yr$^{-1}$
(e.g., \citealt{zfg13}).

Our parallaxes likewise agree extremely well with \Hipp\/ for two of our
targets, but our statistical uncertainties are about 2.5--4 times smaller. For
HD\,84937, our FGS parallax measurement is smaller than that found by \Hipp,
by about 2$\sigma$ in units of the \Hipp\/ uncertainty.  We note that offsets
of a similar amount and sign between \Hipp\/ parallaxes and precise FGS and/or
radio-interferometric parallaxes have sometimes been found by other
investigators, e.g., for the Pleiades cluster (\citealt{snb05}; \citealt{mrm13}).

\section{Atmospheric Parameters and Chemical Abundances}
\label{sec:atm}

In this section we derive atmospheric parameters and the abundances of those
elements that are of particular relevance for the determination of the ages of
metal-poor stars from fits of isochrones.  Most important are O and Fe, but we
also comment on the abundances of C and N as well as the $\alpha$-capture
elements, Mg, Si, and Ca.  [We include HD\,19445 among the stars that are
subjected to a spectroscopic analysis because it is used in \S~\ref{sec:ages}
to check the reliability of isochrones for [Fe/H] $\sim -2$ at $M_V \approx 5$
($\sim 1$ mag below the turnoff).  HD\,19445 is the best available subdwarf
for this purpose because of its large and well-determined parallax from
{\it Hipparcos} ($\pi = 24.92 \pm 0.91$ mas; \citealt{vl07}).] 

The abundances were derived from equivalent widths (EWs) measured in
high-resolution UVES (\citealt{ddk00}) and HARPS spectra (\citealt{mpq03})
obtained during the ESO programs that are listed in the footnote of
Table~\ref{tab:tab3}.  An exception is HD\,132475 for which the EWs of the
O\,{\sc i} triplet lines were measured from ESO/CES spectra (\citealt{jeg05}).
The UVES and HARPS spectra have resolutions around $10^5$ and signal-to-noise
(S/N) ratios ranging from 400 to 800.  As the continuum in the spectra of these
metal-poor stars is well defined, the high S/N makes it possible to measure
equivalent widths of weak lines with very high accuracy.  A comparison of EWs
measured in overlapping UVES and HARPS spectral regions shows a {\it rms}
dispersion of only 0.6\,m\AA, suggesting that the mean EW is measured
with an accuracy of $\pm 0.3$\,m\AA .

Plane parallel (1D) model atmospheres interpolated to the $\teff$, $\log\,g$,
and [Fe/H] values of the stars were obtained from the $\alpha$-element enhanced
([$\alpha$/Fe] $= +0.4$) MARCS grid (\citealt{gee08}) and used to calculate EWs
as a function of element abundance, assuming local thermodynamic equilibrium
(LTE).  Interpolation to an observed EW then yields the corresponding abundance.
A microturbulence of 1.4\,km\,s$^{-1}$ was adopted. This value is obtained for
HD\,132475 by demanding that the derived Fe abundance shows no systematic
dependence on EW. For the other stars, the lines are so weak that the 
microturbulence parameter has no significant effect on the derived abundances.

\subsection{$\teff$ and $\log\,g$}
\label{subsec:teff}

The effective temperatures given in Table~\ref{tab:tab4} are based on the
calibration of the infrared flux method (IRFM) by \citet{crm10} and taken 
from \citet{mcr10}, except in the case of HD\,140283. For this star, we detect
weak interstellar Na\,{\sc i}\,D lines ($EW = 23.0$ and 11.5\,m\AA ), which 
implies $E(B-V) = 0.004$ according to the calibration of \citet{ama10}.
This increases the IRFM temperature of HD\,140283 from $\teff = 5777$\,K,
as reported by Mel\'{e}ndez et al., who assumed $E(B-V) = 0.0$, to
$\teff = 5797$\,K.  The statistical errors associated with the IRFM temperatures
are on the order of $\pm 60$\,K according to \citet{crm10}, but in addition,
there could be systematic errors of the particular $\teff$-scale that is
adopted.  For instance, using a different implementation of the IRFM based on
2MASS photometry, \citet{ghb09} derived $T_{\rm eff}$ values for HD\,19445,
HD\,84937, and HD\,140283 that are an average of 80\,K cooler than those by
Casagrande et al.  In the case of HD\,132475, they find a higher temperature
(5815\,K), but this is probably because they assume a reddening of $E(B-V) =
0.03$ instead of the value that we favor, $E(B-V) = 0.007$, which is based on
the strength of interstellar Na\,{\sc i}\,D lines. 

As a check of the $\teff$-scale, we have compared the IRFM-based temperatures
with those derived from the wings of the Balmer lines.  This spectroscopic
determination has the advantage of being independent of interstellar
reddening.  For the best studied star, HD\,140283, \citet{gls04} obtained 
$\teff = 5773$\,K from H$\alpha$ and H$\beta$, while \citet{aln06} found 
$\teff = 5753$\,K from H$\alpha$ and \citet{naa07} determined $\teff = 5849$\,K
from H$\beta$.  The average of these values, 5791\,K, agrees well with our
IRFM temperature of 5797\,K, but as discussed by Nissen et al., there may be
systematic errors in the Balmer-line temperatures related to the temperature
structures of model atmospheres, non-LTE effects, and line-broadening theory.
For 10 halo stars having $T_{\rm eff}$ from both H$\alpha$ and H$\beta$,
the H$\beta$ temperatures are systematically higher than those from H$\alpha$
by $\Delta T_{\rm eff} = 64 \pm 28$\,K.  This suggests that the spectroscopic 
$\teff$-scale is uncertain by approximately $\pm 70$\,K.  Furthermore,
for a sample of 30 metal-poor halo stars, \citet{crm10} compared their IRFM
temperatures with those derived from the hydrogen lines by \citet{ber08} and
\citet{naa07}.  As seen in their Fig.~12, there is good agreement between the
two scales at $5600 < \teff\ < 6000$\,K, whereas the IRFM temperatures tend to
be higher by 50 to 100\,K at $\teff > 6100$\,K.  In summary, we conclude that
there is no evidence from the presently available spectroscopic temperatures
of halo stars that the \citet{crm10} IRFM scale is too low, but it may be
50--100\,K too high for metal-poor halo stars in the turnoff region.

Surface gravities are derived from the standard relation
\begin{eqnarray} \log \frac{g}{g_{\odot}}  = 
 \log \frac{\cal{M}}{\cal{M}_{\odot}} +
 4 \log \frac{T_{\rm eff}}{T_{\rm eff,\odot}} +
 0.4 (M_{\rm bol} - M_{{\rm bol},\odot})
\end{eqnarray}
where $\cal{M}$ is the mass of the star and $M_{\rm bol}$ is the absolute
bolometric magnitude (see \S~\ref{sec:ages}).  From the errors, which are
estimated to be $\pm 0.03 \cal{M}_{\odot}$ in mass, $\pm 60$\,K in
$T_{\rm eff}$, and $\pm 0.03$ in the bolometric correction, as well as the
error of $M_{\rm V}$ arising from the parallax uncertainty, we obtain
$\sigma (\log g) \, = \, \pm 0.04$\,dex for HD\,19445 
and $\sigma (\log g) \, = \, \pm 0.03$\,dex for the other stars.

\subsection{O and Fe abundances}
\label{subsec:OFe}

The abundances of oxygen and iron were determined from the lines that are
listed in Table~\ref{tab:tab3}.  The oscillator strengths ($\log\,gf$) of the
O\,{\sc i} triplet lines are from the theoretical work by \citet{hbg91}, while
those adopted for the Fe\,{\sc ii} lines are from \citet{mb09}, who determined
accurate $gf$-values based on lifetime measurements of the atomic energy levels
and calculations of relative line ratios within a given multiplet.  The results
from individual lines clearly agree very well. In the case of oxygen, the
{\it rms} scatter of $A$(O)\,\footnote{For an element X, $A({\rm X}) \equiv
{\rm log} \, (N_{\rm X}/N_{\rm H}) +12.0$.} is on the order of $\pm 0.04$\,dex
and the scatter of $A$(Fe) is $\pm 0.06$\,dex.

Non-LTE corrections for the oxygen triplet lines are adopted from the
calculations of \citet{fab09}, who considered a model atom with 54 energy levels
and adopted electron collision cross sections from \citet{bar07}.  Inelastic
collisions with hydrogen atoms were described by the classical Drawin formula
(\citealt{dra68}), scaled by a factor $S_{\rm H}$. The calculations were
performed for both $S_{\rm H}$ = 0 and 1, which enables us to interpolate the
non-LTE corrections to $S_{\rm H}$ = 0.85, i.e., the value determined by  
\citet{pak09} from a study of the solar centre-to-limb variation of the
O\,{\sc i} triplet lines.  This leads to mean non-LTE corrections for the three
O\,{\sc i} lines ranging from $-0.11$ dex in the case of HD\,19445 to
$-0.19$\,dex in HD\,84937.  By comparison, the non-LTE correction for the Sun
is $-0.20$\,dex. 

The high excitation O\,{\sc i} triplet lines are formed deep
in the stellar atmospheres, where the effects of atmospheric inhomogeneities
on the temperature stucture are small.  According to inhomogeneous models of
metal-poor halo stars, the 3D--1D,\,MARCS correction of oxygen abundances
derived from the O\,{\sc i} triplet is negligible (\citealt{asp05}, his Fig. 8). 

The Fe abundance determined from Fe\,{\sc ii} lines is practically unaffected
by departures from LTE (\citealt{mgs11}; \citealt{lba12}).  There is, however,
a small 3D--1D,MARCS correction of approximately +0.05\,dex in metal-poor stars
(\citealt{npa02}, \citealt{asp05}).

The abundance determinations are summarized in Table~\ref{tab:tab4}. The
statistical errors arising from uncertainties in the EWs and those of the
$\teff$ and $\log\,g$ values are on the order of $\pm 0.07$\,dex for $A$(O) and
$\pm 0.04$\,dex in the case of $A$(Fe). Systematic errors due to uncertainties
in the $gf$-values, the non-LTE effects, and the 3D--1D corrections are larger:
we estimate total uncertainties of $\pm 0.15$\,dex for $A$(O) and $\pm
0.10$\,dex for $A$(Fe).  The table also lists [Fe/H] and [O/Fe] values
corresponding to the solar abundances reported by \citet[who give
$A$(O)$_{\odot} = 8.69$ and $A$(Fe)$_{\odot} = 7.50$]{ags09}. As noted, [O/Fe]
in HD\,84937 is about 0.15\,dex lower than in the other three stars, but given
that the error of the differential values of [O/Fe] is $\pm 0.08$\,dex (mostly
due to the temperature uncertainties) this could be an accidental $2\,\sigma$
deviation.  (Alternatively, the lower abundance may be due to gravitational
settling, which is expected to have bigger effects on the surface abundances of
turnoff stars than those on the MS or SGB; see \citealt{rmr02}.)

The O and Fe abundances agree fairly well with results from recent studies.
In a large survey of 825 stars in the local disk and halo, \citet{ral13}
derived oxygen abundances from the O\,{\sc i} triplet lines using their own
non-LTE calculations. The four stars discussed in this paper are included in
their survey.  The mean differences (this paper $-$ Ram\'{\i}rez) and {\it rms}
deviations are: 
$\Delta \, T_{\rm eff} = 30 \, \pm 35$\,K,
$\Delta \, {\rm log}\,g = -0.02 \, \pm 0.06$,
$\Delta \, {\rm [Fe/H]} = -0.09 \, \pm 0.05$,
$\Delta \, {\rm [O/H]} = -0.05 \, \pm 0.13$, and
$\Delta \, {\rm [O/Fe}] = +0.04 \, \pm 0.12$.
The {\it rms} deviations correspond well to the estimated errors in the two
studies.  There seems, however, to be a significant systematic difference
in the [Fe/H] values in the sense that the Ram\'{\i}rez et al.~metallicity 
scale is about 0.10\,dex higher than ours. On the other hand, our metal 
abundances agree almost exactly ($\Delta {\rm [Fe/H]} = +0.01 \, \pm 0.02$)
with those of \citet{crm10} and \citet{mcr10}, who adopted mean values of
[Fe/H] from a number of recent high-resolution studies.
  
The oxygen abundance may also be determined from the forbidden [O\,{\sc i}]
line at 6300\,\AA , which is, however, very weak in spectra of 
metal-poor F and G dwarf stars.  This line could not be detected in
HD\,19445 and HD\,84937, but in our spectrum for HD\,132475, we measured an
equivalent width of $EW = 1.7$\,m\AA \,(after a small correction
for the contribution of the Ni\,{\sc i} blend (\citealt{ala01}),
and in the case of HD\,140283, \citet{npa02} obtained $EW = 0.5$\,m\AA .
Non-LTE corrections are negligible, but there is a significant 3D--1D
correction on the order of $-0.2$\,dex (see Nissen et al.~2002), because the
[O\,{\sc i}] line is formed in the upper layers of the atmosphere, where the
temperature is lower in 3D models than in 1D models.  Taking this into account,
we derive $A$(O) = 7.60 for HD\,132475 and $A$(O) = 6.95 for  HD\,140283.
In the case of HD\,132475, the oxygen abundance from the [O\,{\sc i}] line
is $\sim \! 0.2$\,dex lower than derived from the triplet, whereas the two
O abundances happen to agree for HD\,140283. As the oxygen
abundances derived from the [O\,{\sc i}] line have statistical errors
on the order of $\pm 0.2$\,dex due to the weakness of this line, we consider 
the results from the O\,{\sc i} triplet to be more reliable.

The [O\,{\sc i}] $\lambda \, 6300$ line is of sufficient strength in metal-poor
K giants to allow a precise determination of their oxygen abundances. In an
extensive study of such stars, \citet{cds04} found a plateau of
[O/Fe] $\simeq 0.7$ at [Fe/H] $< -2$ based on a 1D model atmosphere analyses.
Applying 3D-1D corrections of $-0.1$\,dex (\citealt{cat07}) for giants,
the plateau value decreases to [O/Fe] $\simeq 0.6$, in good agreement with the
average ratio, [O/Fe] $= 0.57$ that has been derived in this paper for the four
dwarfs and subgiants.

\subsection{C, N, and $\alpha$-element Abundances}
\label{subsec:cn}

The abundances of carbon and nitrogen relative to that of iron are not far
from the solar ratios.  From equivalent widths of C\,{\sc i} lines near
9100\,\AA , we get [C/Fe] $= -0.01$ for HD\,84937 and [C/Fe] $= 0.08$ for
HD\,140283 if non-LTE corrections from \citet{fna09} corresponding to 
$S_H = 1$ are adopted.  If $S_H = 0$, these abundance ratios are decreased
by 0.08\,dex.  In the case of HD\,132475, \citet{ncc14}
derive [C/Fe] = 0.15 from the weak C\,{\sc i} lines at 5250 and 5380\,\AA . We
have no data for HD\,19445, but \citet{tt13} obtained [C/Fe] $= +0.03$
from a set of infrared  C\,{\sc i} lines at 1.068--1.069 $\mu$m. Within the 
estimated errors, typically $\pm 0.15$\,dex, all of these values are compatible
with a solar C/Fe ratio. Concerning nitrogen, we note that a model-atmosphere
synthesis by \citet{ier04} of the NH band at 3360\,\AA\ in UVES spectra
yields [N/Fe] $= 0.35$ for  HD\,19445, $< 0.0$ for HD\,84937, and 0.05
for HD\,140283. These values are based on 1D model atmospheres.
According to \citet{asp05}, large negative 3D corrections should be applied,
because the NH band is formed high in the atmospheres, which decreases
[N/Fe] by 0.4 to 0.5 dex. For HD\,132475, we have not been able to find any
determinations of the nitrogen abundance based on high-resolution spectra,
but intermediate-resolution observations of the NH band yield [N/Fe] $\sim
-0.5$ (\citealt{cbk87}). Hence, although the N abundances are uncertain,
it is clear that none of the four stars considered here belong to the rare
class of N-rich halo stars (\citealt{bn82}), and that nitrogen gives a
negligible contribution to the total abundance of the CNO elements.

The abundances of the $\alpha$-capture elements, Mg, Si, and Ca, have been
determined from one Mg\,{\sc i} line ($\lambda 5711.1$), two Si\,{\sc i} lines
($\lambda 6155.1$ and $\lambda 6237.3$), and eight Ca\,{\sc i} lines in the
wavelength range 6100--6440\,\AA .  The sources of the relevant
$gf$ values are \citet{ct90} for Mg, \citet{sgm09} for Si,
and \citet{sr81} for Ca.  Non-LTE corrections of about $+0.15$\,dex
for the Mg\,{\sc i} line were adopted from \citet{zg00}, whereas the
corrections for the Si and Ca lines are vanishingly small 
(\citealt{sgm09}; \citealt{mkp07}). In addition, we include small
3D--1D corrections from \citet{asp05}.  Based on all of this, we derive the
[$\alpha$/Fe] values given in Table~\ref{tab:tab4} (adopting solar abundances
from \citealt{ags09}).  As shown in this table, the [$\alpha$/Fe] values of
HD\,19445, HD\,84937, and HD\,132475 are close to 0.40\,dex, while that for
HD\,140283 (0.26\,dex) is lower. This may be a statistical fluctuation
because the error of [$\alpha$/Fe] is on the order of $\pm 0.1$\,dex.
Alternatively, HD\,140283 could belong to the population of "low-$\alpha$"
halo stars discovered by \citet{ns10}, although it would then be puzzling why
HD\,140283 does not have a lower value of [O/Fe] than the other stars.  We
note in this connection that the possible low [$\alpha$/Fe] has no influence on
the abundance determinations: if a MARCS model with [$\alpha$/Fe] $= +0.26$ is
applied instead of the canonical value of $+0.40$\,dex, there is no significant
change of the derived abundances.

\subsection{Initial Chemical Abundances}
\label{subsec:initial}

The abundances described in the previous subsections refer to the present
compositions of the stellar atmospheres.  As discussed by \citet{bnv13}, the
observed surface abundances should be adjusted for the effects of diffusive
processes in order to yield the bulk composition of the stellar 
interior.  (If gravitational settling is not treated, the age of HD\,140283
would exceed the age of the universe as determined from {\it Planck} and {\it
WMAP} observations by $\gta 1.5$ Gyr.)  However, it is clear from the studies by
\citet{lkb08}; \citet[and references therein]{nkr12}, and \citet{gkr13} that
diffusive stellar models are not able to reproduce the observed variations of
the metal abundances of stars lying between the TO and the base of the
giant branch in GC CMDs unless additional mixing (below envelope convection
zones) is invoked to limit the efficiency of diffusion in their surface layers
(also see \citealt{rmr02}). In the case of NGC\,6397, which has [Fe/H]
$\approx -2.0$ (e.g., \citealt{cbg09}), Nordlander et al.~concluded that the
difference between the initial and observed metal abundances in the case of
cluster subgiants with $\teff \approx 5800$~K (similar to that of
HD\,140283) is 0.1--0.15 dex, depending on how this extra mixing is treated.
Similar, or slightly smaller, differences were obtained by Gruyters et al.~in
their investigation of NGC\,6752 ([Fe/H] $\approx -1.55$).

Based on these findings, we decided to correct the [Fe/H] value that we
determined for HD\,132475 by $+0.10$ dex and those of the other three stars
by $+0.15$ dex.  (As mentioned in \S~\ref{subsec:OFe}, a larger adjustment
could well be more appropriate for HD\,84937.)  If the same corrections are
applied to the absolute abundances of all of the metals, the inferred [$m$/Fe]
values remain unaffected.  However, since the ages of very metal-poor subgiants,
at their observed absolute magnitudes, depend primarily on their absolute
oxygen abundances (see below), the adoption of increased [O/H] values by
0.1--0.15 dex does have the effect of reducing the ages derived for them by
0.3--0.5 Gyr, which is not negligible.  That is, the ages of HD\,132475 and
HD\,140283 that are derived in this study would have been higher by up to 0.5
Gyr if the aforementioned adjustments to the observed abundances had not been
made.

\section{Stellar Evolutionary Models}
\label{sec:models}

For the comparisons with observations presented below, stellar models have been
computed using the version of the Victoria stellar evolution code that has been
described in considerable detail by \citet{vbd12}.  The main differences
relative to previous versions (e.g., see \citealt{vsr00}) are the incorporation
of the latest rates for the H-burning reactions and a careful treatment of the
gravitational settling of helium, along with the implementation of additional
mixing below envelope convection zones (when they are present) in order to
satisfy the solar and ``Spite plateau" (\citealt{ss82}) lithium abundance
constraints (e.g., \citealt{rmr02}).\footnote{Metals diffusion (aside from Li)
is not treated in the Victoria code; consequently, we are unable to compare
the predicted surface abundances of our models with those observed.  However,
insofar as isochrones are concerned, this physics mainly affects the $\teff$
scale in the vicinity of the TO.  At the locations of HD\,132475 and
HD\,140283 near the middle of the SGB, the differences between isochrones
that allow for the settling of helium, on the one hand, and those that also
treat the diffusion of the metals, on the other, will be barely discernible,
judging from the comparisons of evolutionary tracks presented by VandenBerg et
al.~(2012, their Fig.~1).  Only in the case of HD\,84937 will the neglect of
metals diffusion be of some concern.}  Interestingly, the net effect of these
improvements on the TO luminosity versus age relations that are applicable to
old Population II stars is quite small: the $\sim 10$\% reduction of the TO age
that occurs when diffusive processes are treated (\citealt{pv91}) is largely
offset by a comparable increase of this quantity when recent determinations of
the rate of the critical $^{14}$N$(p,\,\gamma)^{15}$O reaction (\citealt{fic04},
\citealt{mfg08}) are adopted in computations of stellar models (as first
reported by \citealt{icf04}).

The physics improvements that have been made have much bigger consequences for
the predicted temperatures of stars and the morphologies of their evolutionary
tracks than for the dependence of the TO luminosity on age.  However, the model
effective temperature ($\teff$) scale is subject to many uncertainties
(including, in particular, the treatment of convection and the atmospheric
boundary condition), and its accuracy can be assessed only through comparisons
with empirical determinations.  As shown by \citet[see their Fig.~10]{vcs10},
current Victoria-Regina isochrones reproduce the locations of field halo
subdwarfs that have well determined distances from {\it Hipparcos}
(\citealt{vl07}) very well if the temperatures of the latter are close to the
values derived by \citet{crm10} from their calibration of the infrared-flux
method (IRFM).  In fact, similar success is obtained on many CMDs (also see
\citealt{vbf14}) if color--$\teff$ relations (\citealt{cv14}) based on the
latest MARCS model atmospheres (\citealt{gee08}) are used to transpose the
models to the observational planes.  (The same transformations are used in
this study to derive bolometric corrections and colors.)

Moreover, \citet{vbl13} have shown that the same isochrones provide a close
match to the shapes of GC CMDs from at least $\sim 2$ mag below the TO through
to approximately the middle of the subgiant branch (SGB), independently of the
assumed age and metallicity (see their Figs.~2 and 3).  Indeed, this provides
an argument that the errors of the predicted TO temperatures due to the neglect
of metals diffusion (see footnote 7) are probably quite small.  The main
{\it concern} with these models is that the predicted locations of cluster RGBs
appear to be somewhat too cool/red, but the primary cause of this discrepancy
has not yet been identified.  It is possible, for instance, that the assumption
of a constant value of the mixing-length parameter, $\amlt$, is at the root of
this difficulty, given that calibrations of this parameter using 3D model
atmospheres suggest that it should vary somewhat with $\teff$, $\log\,g$, and
[Fe/H] (\citealt{ts11}; \citealt{mwa14}).  Errors in the color transformations
and the treatment of the surface boundary condition could also be contributing
factors.  Indeed, we suspect that it will be difficult to resolve this issue as
long as the uncertainties associated with the derived temperatures and
metallicities of metal-poor giants (in the field or in star clusters) remain as
large as they are (typically $\sim \pm 70$~K and $\sim \pm 0.1$--0.2 dex,
respectively).  Worth mentioning is the fact that the aforementioned problem
concerning the RGB, which has little or no impact on the ages derived in this
study, is not unique to models that are generated using the Victoria code.  As
shown by \citet{vbd12}, evolutionary tracks from the zero-age main sequence to
the RGB tip (and such properties as the helium core mass at core helium
ignition) are in excellent agreement with those obtained using the completely
independent, very versatile MESA program (\citealt{pbd11}), when essentially
the same up-to-date physics is assumed.
 
We note, finally, that new grids of tracks and isochrones (which will be
the subject of a separate paper by D.~VandenBerg et al.~2014, in preparation)
are used in the following analysis.  Whereas the models used by \citet{bnv13}
to derive our initial estimate of the age of HD\,140283 assumed the solar mix
of the metals given by \citet{gs98}, with various enhancements to the abundances
of the $\alpha$-elements (O, Ne, Mg, etc.) and then scaled to the observed
[Fe/H] value, the newly computed models assume the solar metal abundances given
by \citet{ags09} as the base mixture.  At low metallicities, the abundances of
all of the $\alpha$-elements, except oxygen, are assumed to be enhanced by the
fixed amount of 0.4 dex (i.e., [$m$/Fe] $= 0.4$).  However, because TO
luminosity versus age relations are such a strong function of the oxygen
abundance, which is subject to considerable uncertainties and which appears to
vary to some extent from star-to-star (e.g., \citealt{fna09}; \citealt{rmc12})
and possibly from cluster-to-cluster, separate grids of models have been
computed for [O/Fe] $= 0.4, 0.6, 0.8$ and 1.0.  Moreover, to address possible
helium abundance variations, grids of evolutionary tracks have also been
computed for helium mass-fraction abundances $Y = 0.245$ and 0.285 (for each
assumed metallicity).  Fully consistent low- and high-temperature opacities
for each chemical mixture have been used in the model computations.  

As described by \citet{vbf14}, the software developed by P.~Bergbusch permits
simultaneous interpolations in three chemical abundance parameters ([Fe/H], $Y$,
and either [$\alpha$/Fe], if the abundances of all of the $\alpha$-elements vary
together, or [O/Fe], if a constant [$m$/Fe] value is assumed for the heavier
$\alpha$-elements).  As a result, we have the capability to generate isochrones
for very close to the preferred chemical abundances.  While there will still be
small differences between the assumed and actual [$m$/Fe] values for many of the
metals, the effects of such differences on isochrones that are applicable to
low-metallicity stars are of little consequence (see below).  What is important
is that we are able to treat [Fe/H], [O/Fe], and $Y$ as free parameters.  (At
this time, the model grids span the range in [Fe/H] from $-1.4$ to $-2.8$,
though they will be extended to both lower and higher iron abundances in the
coming months.)

%Insofar as the calculation of isochrones is concerned (see the Appendix), the
%neglect of metals diffusion is not a serious omission because this physics
%mainly affects the predicted temperatures of stars (in a relatively minor way.

\section{The Ages of the Three Subgiants}
\label{sec:ages}

\begin{figure}
\plotone{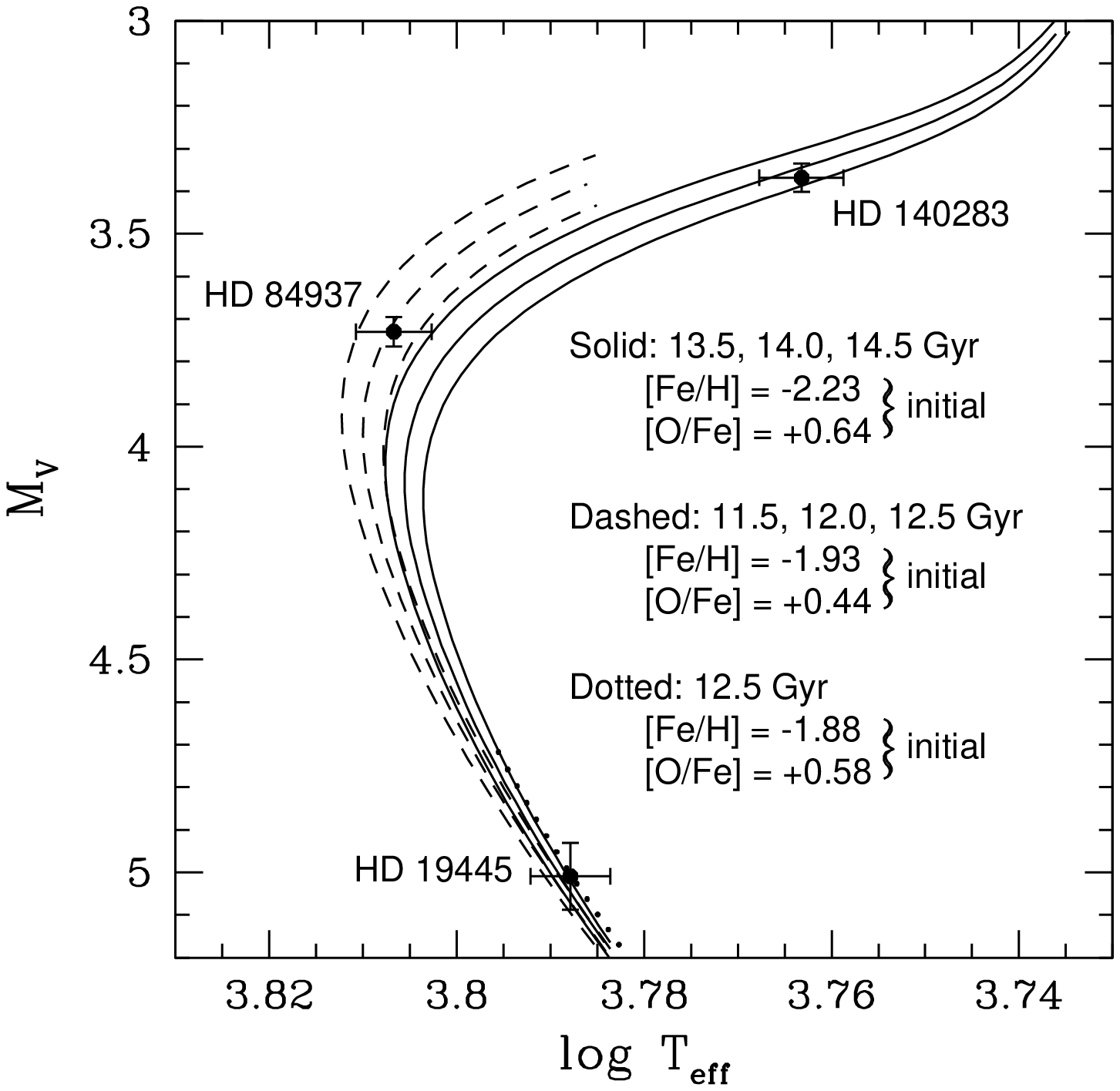}
\caption{Overlay of isochrones for the indicated ages and {\it initial} chemical
abundances onto the locations of HD\,19445, HD\,84937, and HD\,140283 on 
the $(\log\,\teff,\,M_V)$-diagram.}
\label{fig:fig1}
\end{figure}

Figure~\ref{fig:fig1} compares isochrones for the indicated ages and {\it
initial} chemical abundances with the observed locations of HD\,84937 and
HD\,140283 on the $(\log\teff,\,M_V)$-diagram.  The adopted $V$ magnitudes,
reddenings, and the values of $M_V$ that correspond to the parallaxes listed
in Table \ref{tab:tab2}, suitably adjusted by $A_V = 3.07\,E(B-V)$ (e.g.,
\citealt{mcc04}), are given in Table~\ref{tab:tab5}, which also lists the
observed Johnson-Cousins $B-V$ and $V-I$ colors from \citet{crm10}.
Irrespective of whether or not the statistical Lutz-Kelker corrections
(\citealt{lk73}) {\it should} be applied,\footnote{According to \citet[see
p.~209]{per09}, ``the parallax of an individual star is not itself biased, and
bias correction for an individual star outside of the context of a
parallax-limited sample is not appropriate".  However, others (e.g.,
\citealt{bmn09}) have argued that Lutz-Kelker corrections should be applied when
calculating the absolute magnitudes of single stars.} they are negligible for
the subgiants in our sample because of the high precision of our parallax
determinations ($\sigma_\pi/\pi \lta 0.02$).  For instance, using an equation
provided by \citet{han79}, these corrections amount to only $-0.0023$ mag or
less (in an absolute sense).  Consequently, they have not been included in the
tabulated absolute magnitudes.

The main reason why HD\,19445 has been included in this plot is to demonstrate
that isochrones satisfy the $\teff$\ constraint provided by this star very well
(as in the case of local subdwarfs of higher metallicity; see VandenBerg et
al.~2010, 2014).  For the sake of clarity, only a small segment of a 12.5 Gyr
isochrone for the metallicity of HD\,19445 has been plotted.  Clearly, an
appreciably older or younger isochrone would have served our purposes just as
well given the fairly large uncertainties in the properties of HD\,19445 and
the small separations between isochrones of different age at $M_V \gta 5$.
Faint subdwarfs provide good tests of the predictions of stellar models and
they represent one of the favored standard candles for distance determinations
precisely because their properties are nearly independent of age.

HD\,84937 appears to be just beginning its SGB evolution, and in fact, it is
more correctly described as a TO star.  At the location of HD\,84937,
isochrones for its observed metal abundances are more vertical than horizontal;
consequently, the uncertainty of its age is due mostly to the uncertainty
associated with its $\teff$\ rather than that of its $M_V$.  According to the
interpolation results given in Table~\ref{tab:tab6}, the age of HD\,84937 is
12.09 Gyr, with uncertainties of $\pm 0.14$ Gyr and $\pm 0.63$ Gyr from the
vertical and horizontal error bars, respectively.  The actual uncertainty will
be larger than this since, in particular, the model $\teff$ scale is uncertain
by at least $\sigma\,(\log\teff) \sim 0.005$, which is the horizontal shift in
the isochrones that is permitted by the $1\,\sigma$ error bars of HD\,19445.
(A further complication is the possibility that we have underestimated
the effects of gravitational settling on the surface abundances of HD\,84937,
which would also impact the inferred age.)  By comparison, the {\it Hipparcos}
parallax, $\pi = 13.74 \pm 0.78$ mas (\citealt{vl07}), which implies $M_V =
3.98 \pm 0.12$, would have placed this star right at the turnoff of the 12.5
Gyr isochrone for its metallicity.  (Note that the tabulated masses were used
to calculate the gravities which are listed in Table~\ref{tab:tab4} using
equation 1.) 

\begin{figure}
\plotone{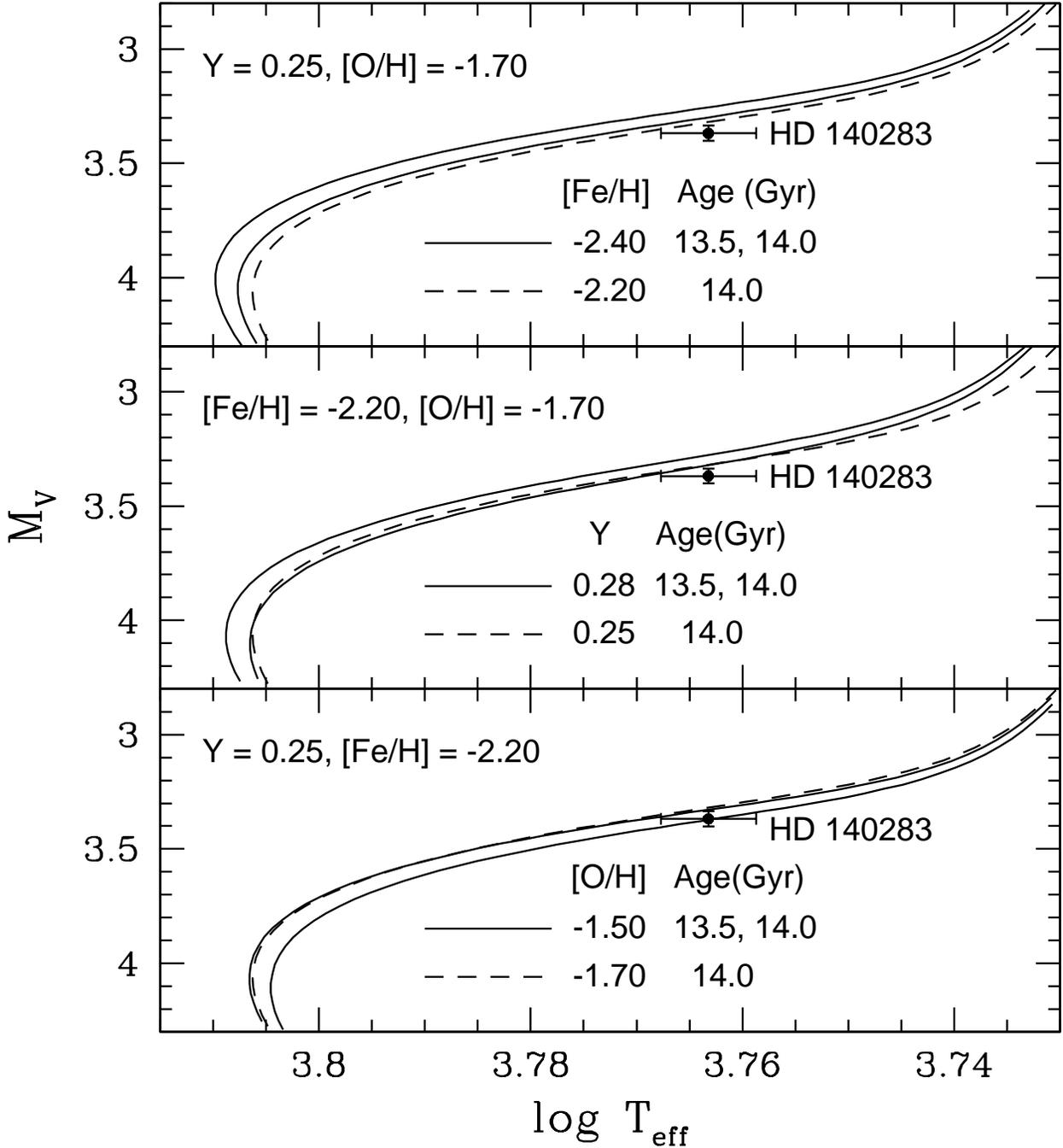}
\caption{Plot to illustrate the differences between 14.0 Gyr isochrones in which
only the values of [Fe/H], $Y$, and [O/H] are varied in turn (upper, middle,
and lower panels, respectively).  For the cases represented as {\it solid
curves}, isochrones for two ages (13.5 and 14.0 Gyr) have been plotted so that
the age difference implied by the vertical separations between the two 14.0 Gyr
isochrones in each panel, at the color of HD\,140283, can be estimated.  Ages
at low metallicities are much more dependent on [O/H] than [Fe/H] or $Y$.}  
\label{fig:fig2}
\end{figure}

Similar interpolations in isochrones computed for the measured abundances of
HD\,140283 yield an age of 14.27 Gyr, with the vertical and horizontal error
bars implying age uncertainties of $\pm 0.38$ Gyr and $\pm 0.37$ Gyr,
respectively (see Table~\ref{tab:tab6}).  Because the SGB has a relatively
shallow slope, the consequences of the $\teff$\ uncertainty for the inferred
age is much less of a concern than in the case of HD\,84937, and indeed,
the parallax (and hence $M_V$) uncertainty is the larger of these two
contributors to the error budget by just a small amount.  Thus, despite carrying
out a new spectroscopic study of HD\,140283 and using new sets of stellar
models, the age of this star differs only slightly from the determination of
$14.46 \pm 0.31$ Gyr reported by \citet{bnv13}.  This is not too surprising,
however, since the values of [Fe/H] and [O/Fe] which have been derived in this 
investigation are quite similar to those adopted by Bond et al., and it is to
be expected that the effects of small metal abundance differences in the
respective isochrones will be of little consequence as long as nearly the same
{\it absolute} oxygen abundance is assumed.  This claim is supported by the
isochrone comparisons presented in Figure~\ref{fig:fig2}.  

In the uppermost panel of Fig.~\ref{fig:fig2}, the models assume the same helium
and oxygen abundances ($Y = 0.25$, [O/H] $= -1.70$), but different metallicities
by $\delta\,$[Fe/H] $ = 0.2$ dex (and hence different values of [O/Fe] by the
same amount).  Since the [Fe/H] value represents the factor used to scale the
$\log N$ abundances of all of the metals from the assumed solar mixture (with
the adopted $\alpha$-element abundance enhancements taken into account) to the
metallicity of interest, {\it all} of the metals other than oxygen have higher
absolute abundances by 0.2 dex in the [Fe/H] $= -2.2$ isochrones than in those
for [Fe/H] $= -2.4$.  Despite this difference, the {\it solid} and {\it dashed}
isochrones for an age of 14.0 Gyr have virtually identical TO luminosities;
i.e., with a suitable horizontal shift of either of these two isochrones, they
would overlay each other nearly exactly in the vicinity of the turnoff (and
elsewhere, in fact).  However, in this study, temperature differences are
important, and the slight shift to cooler temperatures that is predicted to
occur when the abundances of the metals are increased will have some 
consequences for the inferred age of HD\,140283.  Judging from the separation
between the 13.5 and 14.0 Gyr isochrones that have been plotted as {\it solid
curves}, a 0.2 dex error in {\it all} of the derived metal abundances (at
constant [O/Fe]) would change the derived age by about 0.2 Gyr: the lower the 
[Fe/H] value, the higher the age.  (We chose not to plot isochrones for the
specific initial abundances and age that we have determined for this star, as
the clarity of the figure was improved by having the model loci slightly offset
from the position of HD\,140283.)

The middle panel of Fig.~\ref{fig:fig2} demonstrates that helium abundance
uncertainties have no significant impact on the age of HD\,140283 due to its
fortuitous location close to the center of the SGB, where isochrones of the same
age and metal abundances, but different $Y$, are virtually coincident.  It is a
well-known result (e.g., see \citealt{vcs12}; \citealt{vbd12}) that, at least at
low metallicities, helium has only a small effect on the slope of the subgiant
branch (at a fixed age), but not on the luminosity near the middle of the SGB.
The fact that the helium abundance does not play a role in the ages of
HD\,140283 thus removes from consideration a parameter ($Y$) that is very
difficult to measure in Population II stars, and which is known to vary within
some GCs (e.g., NGC\,2808; see \citealt{pba07}), possibly from
cluster-to-cluster, and perhaps among field halo stars as well.

In the bottom panel, the strong dependence of age on the oxygen abundance is
illustrated.  In this comparison, the isochrones have the same absolute
abundances of all of the metals, except oxygen, which is varied by
$\delta\,$[O/H] $= 0.2$ dex.  The case represented by the {\it solid} curves
assumes [Fe/H] and [O/Fe] values which differ from those that we have derived
for HD\,140283 by only $+0.03$ and $+0.09$ dex, respectively.  With these
slight upward adjustments, the implied age of HD\,140283 would decrease to
14.0 Gyr and an age of $\sim 13.5$--13.6 Gyr would be contained within the
$1\,\sigma$ error bars.  It is certainly well within the realm of possibility
that the model temperatures are somewhat too high, which may ultimately turn
out to be the main reason why our current best estimate of the age of this star
is greater than the age of the universe by $\sim 0.4$ Gyr.  At this point
in time, we simply do not have any compelling, independent evidence that the 
isochrones are too hot.  Regardless, HD\,140283 is clearly a very old star
that must have formed soon after the Big Bang.

\begin{figure}
\plotone{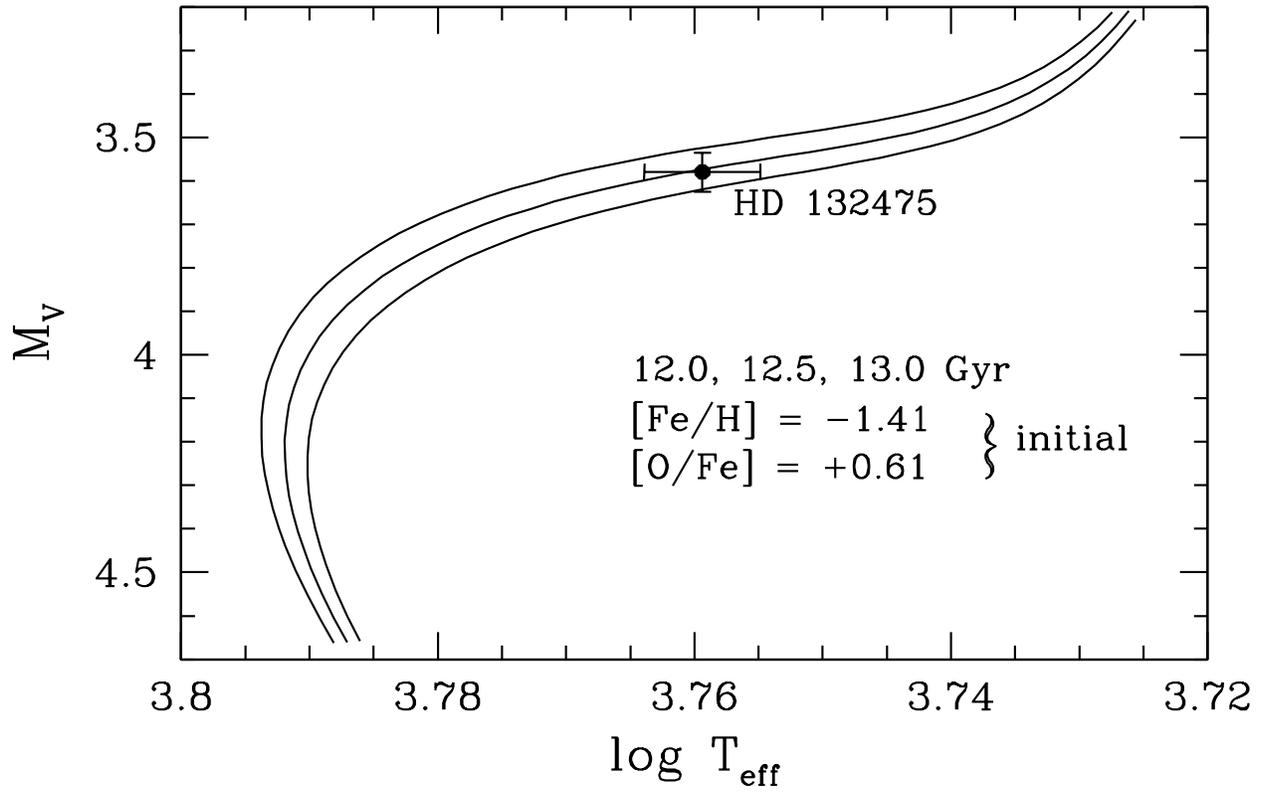}
\caption{Similar to Fig.~1; in this case, the 12--13 Gyr isochrones that have
been plotted are applicable to HD\,132475.}
\label{fig:fig3}
\end{figure}

The age of the last of the three subgiants in our sample, HD\,132475, for
which we have derived {\it initial} abundances corresponding to [Fe/H] $=
-1.41$ and [O/Fe] $= +0.61$, is $12.56 \pm 0.46$ Gyr (see Figure~\ref{fig:fig3}
and Table~\ref{tab:tab6}).  This error bar takes into account only the
uncertainties associated with the FGS parallax.  Because the SGBs of isochrones
for a fixed age have shallower slopes as the metallicity increases, the
ramification of the adopted $\pm 60$~K $\teff$\ uncertainty for the age of
HD\,132475 ($\pm 0.26$ Gyr, which is less than that due to the $M_V$
uncertainty by a factor of 1.8) is considerably smaller than in the case of
HD\,140283.  These two stars apparently differ in age by $\gta 1.5$ Gyr.

As discussed by \citet[see their Table 1]{bnv13}, reasonable estimates of the
uncertainties associated with all of the many factors that play a role in the 
determination of stellar ages (including, e.g., the adopted $V$ magnitudes,  
bolometric corrections, reddenings, and chemical abundances) would imply that 
the net $1\,\sigma$ error bar that should be attached to our age
determinations is $\approx 0.8$ Gyr (or a little higher in the case of
HD\,84937).  Although it seems unlikely that the physics of stellar interiors
will undergo significant revisions in the coming years, since opacities (due
mostly to bound-free and free-free processes involving H and He at low
metallicities) and nuclear reaction rates appear to be well determined,
improvements to the treatment of convection and the atmospheric boundary
condition, among other things, could easily impact the model temperature scale
at the level of $\delta\log\teff = \pm 0.005$--0.01.  Because such variations
are comparable to current uncertainties in the empirical $\teff$ scale,
especially if errors in [Fe/H] are also taken into account, it will be difficult
to advance our understanding of this aspect of stellar models until the
observational constraints are tightened considerably.

\section{Implications of the Three Subgiants for Our Understanding of GCs}
\label{sec:gcs}

In principle, HD\,140283 and HD\,132475 should be very good standard
candles in view of their well-determined $M_V$ values.  (HD\,84937 is less
useful for this purpose because it lies so close to the turnoff.)  However,
because they are evolved stars, the distances of objects that are derived from
them will be accurate only if they share a common age and chemical composition.
Nevertheless, this was believed to be the case when, e.g., \citet[also see
\citealt{pmt98}]{vrm02} obtained $(m-M)_V = 14.62$ for M\,92 by fitting the
cluster subgiants to HD\,140283, on the assumption of $M_V = 3.32 \pm 0.12$
(as determined from the initial {\it Hipparcos} results by \citealt{cgc00}).
(Pont et al.~obtained a slightly larger apparent modulus for M\,92, 14.67 mag,
simply because they assumed that the reddening of HD\,140283 was $E(B-V) =
0.04$ mag, instead of 0.024 mag, as favored by Carretta et al.)  Given the
similarity of the measured [Fe/H] values of M\,92 and HD\,140283 and the
widely held belief that the most metal deficient stars were likely to be
coeval, whether found in the field or in clusters, it was not unreasonable to
obtain an estimate of the distance (and age) of M\,92 in this way.

\begin{figure}
\plotone{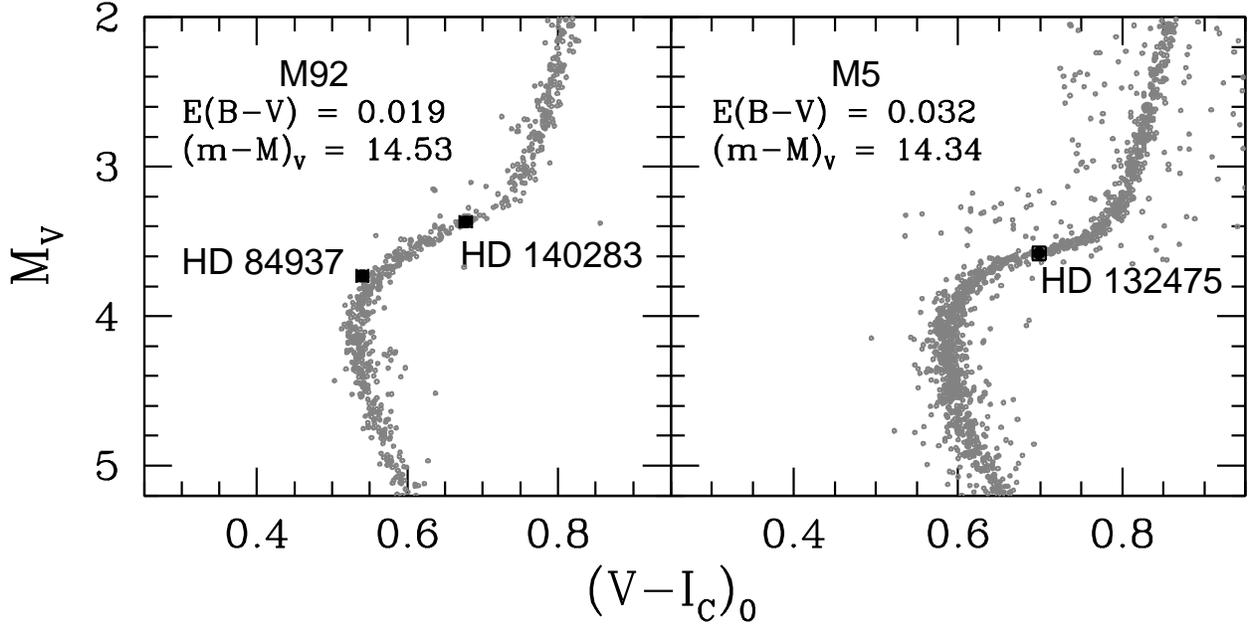}
\caption{Fits of the M\,92 and M\,5 CMDs to HD\,140283 and HD\,132475,
respectively, on the assumption that the field subgiants have the same ages and
chemical abundances as their counterparts in the two globular clusters (see the
text for the source of the photometric data).  If reddenings from \citet{sf11}
are adopted, the derived apparent distance moduli of the two GCs have the values
indicated in each panel.  (HD\,84937, which is more metal-rich than M\,92,
has been included in the left-hand panel just to illustrate where it is located
relative to M\,92 on the $[(V-I_C)_0,\,M_V]$ diagram.  It has not been used
to determine the distance modulus of M\,92.)}
\label{fig:fig4}
\end{figure}

In Figure~\ref{fig:fig4}, we have repeated that exercise, but using our values
of $M_V$ for HD\,140283 and HD\,132475 to determine the distances to M\,92 and
M\,5, respectively.  The indicated values of $(m-M)_V$ correspond to the amounts
that must be subtracted from the apparent magnitudes of the GC stars in order
that the cluster subgiants have the same absolute magnitudes, at the same
intrinsic colors, as the respective field halo subgiants.  In the left-hand
panel, the M\,92 CMD has been fitted to the $M_V$ of HD\,140283 that is
obtained from our FGS parallax ($\pi = 17.18 \pm 0.26$ mas) --- which actually
agrees very well with the values from both the original {\it Hipparcos}
catalogue ($\pi = 17.44 \pm 0.97$ mas, \citealt{plk97}) and the subsequent
reanalysis of the observations ($\pi = 17.16 \pm 0.68$ mas; \citealt{vl07}) ---
together with the current best estimate of its reddening ($E(B-V) = 0.004$; see
\S~2).  This procedure yielded a value of $m-M)_V = 14.53$ for M\,92.  (Note
that HD\,84937 played no role in this determination.  It has been included in
this plot just to show where this star, which is more metal rich than M\,92 by
$\delta$[Fe/H $\gta 0.3$ dex, is located relative to the cluster turnoff.)  The
Johnson-Cousins photometry of HD\,140283 (and HD\,84937) is from \citet{crm10}
while the M\,92 observations were obtained from the ``Photometric Standard
Fields'' archive made publicly available by
P.~B.~Stetson\footnote{http://www2.cadc-ccda.hia-iha.nrc-cnrc.gc.ca/community/STETSON}.
These data have been dereddened assuming $E(B-V) = 0.019$ (from SF11) and
$E(V-I) = 1.3\,E(B-V)$ (e.g., \citealt{dwc78}).  

As there is considerable support for M\,92 having [Fe/H] $\approx -2.2$
(e.g., \citealt{zw84}, \citealt{cg97}), its subgiants should have nearly the
same $M_V$ as HD\,140283, at the same color, provided that they have
approximately the same oxygen abundance and age.  (Note that the effects of
diffusion on the surface chemistry of stars are largely erased by the deepening
convection that occurs along the RGB; consequently, the spectroscopic abundances
that are determined for GC giants will be close to their initial abundances ---
which are the relevant ones for our comparison.   That is, cluster SGB stars
that initially had [Fe/H] $\approx -2.20$ presumably have the {\it observed}
surface metallicity of HD\,140283 at the same evolutionary state, if 
diffusive processes operate at the same rates in both.)  However, a value of
$(m-M)_V$ as small as 14.53 is in conflict with most estimates.  For instance,
\citet{vbl13} recently derived $(m-M)_0 = 14.66$ for the true distance modulus
of M\,92 based on a fit of a zero-age horizontal branch (ZAHB) to the lower
bound of the distribution of its HB stars.  The ZAHBs presented in that study
were shown to satisfy current empirical constraints on RR Lyrae luminosities
(see their Fig.~10) quite well.  Obviously, something is awry and, in the next 
subsection, we try to resolve this dilemma.

Before doing that, a few remarks are in order concerning the right-hand panel
of Fig.~\ref{fig:fig4}.  Our estimate of the {\it initial} iron abundance of
HD\,132475 ([Fe/H] $= -1.41$) is well within the uncertainties of many
determinations of the metallicity of M\,5 (e.g., \citealt{zw84},
\citealt{cbg09}).  Furthermore, [O/Fe] $= 0.4$--0.6 is generally found in GC
stars that constitute the high oxygen end of the O--Na anticorrelation
(including members of M\,5; see \citealt{cbg10}, \citealt{smh13}).  Since the
SGB of M\,5 has little vertical scatter, the total C$+$N$+$O abundance must
be nearly constant, in which case, the low [O/Fe] values that are also found in
most clusters arose from deep mixing along the upper RGB (e.g., \citealt{dv03})
and/or the gas out of which the current stars formed had previously undergone
different amounts of CNO-processing, perhaps in the AGB stars from a slightly
earlier stellar generation (e.g., \citealt{gcb12}) or in a primordial
supermassive star (\citealt{dh14}).  In any case, M\,5 appears to be
chemically quite similar to HD\,132475, and if the latter
is used to determine the distance modulus of the former, one obtains $(m-M)_V =
14.34$.  (As in the case of M\,92, Stetson's ``Photometric Standard Fields"
photometry of M\,5 has been plotted, the reddening is from SF11, and the
source of the Johnson-Cousins $VI$ data for HD\,132475 is \citealt{crm10}.)

Assuming the standard value of $A_V = 3.07\,E(B-V)$, the
true distance modulus of M\,5 is 14.24, which agrees quite well with the
value of $(m-M)_0 = 14.27$ that was obtained by \citet{vbl13} from a fit of a
ZAHB to the cluster horizontal branch stars.  However, the latter adopted 
[Fe/H] $= -1.33$ (\citealt{cbg09}) for M\,5: the assumption of [Fe/H] $=
-1.41$ would have resulted in a larger modulus by a few hundredths of magnitude.
Still, the two independent ways of deriving the distance yield reasonably
consistent results, in contrast with our findings in the case of M\,92, when
HD\,140283 is used as a standard candle.  We now turn to an investigation of
this problem, though we will return to a consideration of M\,5 afterwards.

\subsection{The Distance and Age of M\,92}
\label{subsec:m92}

\begin{figure}
\plotone{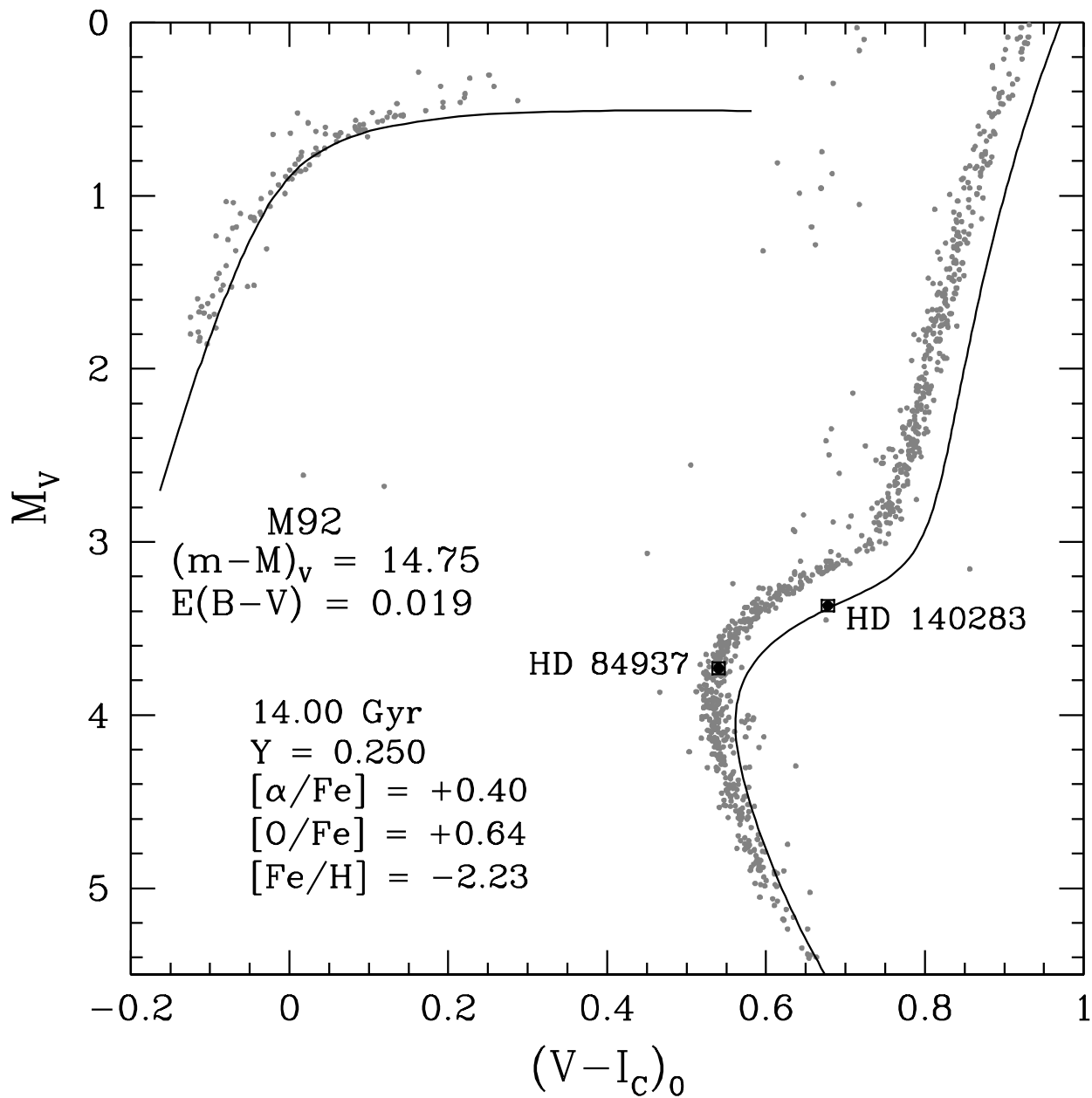}
\caption{Similar to the left-hand panel in the previous figure; in this case,
the apparent distance modulus of M\,92 is based on a fit of a ZAHB to the
lower bound of the distribution of cluster HB stars.  If the higher value of
$(m-M)_V$ is the more accurate determination, HD\,140283 is much fainter than
cluster subgiants that have the same intrinsic $V-I_C$ color.  A 14.0 Gyr
isochrone for the chemical abundances of HD\,140283 has also been plotted.}
\label{fig:fig5}
\end{figure}

Figure~\ref{fig:fig5} plots the same 14.0 Gyr isochrone that appeared in
Fig.~\ref{fig:fig1}, but on the $[(V-I)_0,\,M_V]$-plane.  On this CMD, the
inferred age of HD\,140283 is about 0.30 Gyr younger than our best estimate
from the $(\log\teff,\,M_V)$-diagram, though virtually the same age that was
derived in \S~\ref{sec:ages} is implied by both the $B-V$ and $V-K_S$ colors
of HD\,140283 (not shown).  As discussed by \citet{cv14}, whose color--$\teff$
relations are used throughout this investigation, uncertainties in the
zero-points of the transformations at the level of 0.01--0.015 mag are easily
possible (for any color index).  Alternatively, it may be the observed $V-I$
color of HD\,140283 that is anomalously too blue by a small amount (for
whatever reason).  This is a moot point.  What {\it is} troubling is that the
fit of a fully consistent ZAHB to the cluster counterpart in M\,92 yields an
apparent distance modulus, $(m-M)_V = 14.75$, that is 0.22 mag higher than the
value which is obtained by matching the cluster CMD to HD\,140283 (see the
previous figure).  If the shorter modulus is adopted, not only would the 
cluster HB stars be fainter than our ZAHB models (which, as already mentioned,
do a good job of satisfying empirical RR Lyrae luminosity constraints; see
\citealt{vbl13}), but the cluster main sequence (MS) would be much fainter, at
a given color, than the MS portion of the isochrone for [Fe/H] $ = -2.23$ that
has been plotted in Fig.~\ref{fig:fig5}.  This is problematic because the
separations between isochrones for low metallicities are predicted to be quite
small at $M_V \gta 5$ on the $V-I,\,V$ plane (see VandenBerg et al.~2010, their
Fig.~11).  

These difficulties are almost certainly telling us that M\,92 and HD\,140283
cannot have the same age --- the field halo subgiant must be older --- and
perhaps even that they are chemically dissimilar.  To examine the implications
of possible age and chemical abundance differences, we have opted to fit
isochrones and ZAHB loci to the $F606W,\,F814W$ photometry that was obtained by
\citet{sbc07} using the ACS (Advanced Camera for Surveys) on the {\it Hubble
Space Telescope}.  (The same data were recently used in the determination of the
ages of 55 of the GCs in their sample by \citealt{vbl13}.)  The main reason for
this choice is that a deeper and tighter CMD for M\,92 can be derived from
this archive\footnote{http://archive.stsci.edu/prepds/acsggct} than from
any other publicly available source.  In addition, since we intend to carry
out similar fits of stellar models to the CMD of M\,5, it is an important
advantage of using the Sarajedini et al.~observations that they represent a very
homogeneous data set.
 
\begin{figure}
\plotone{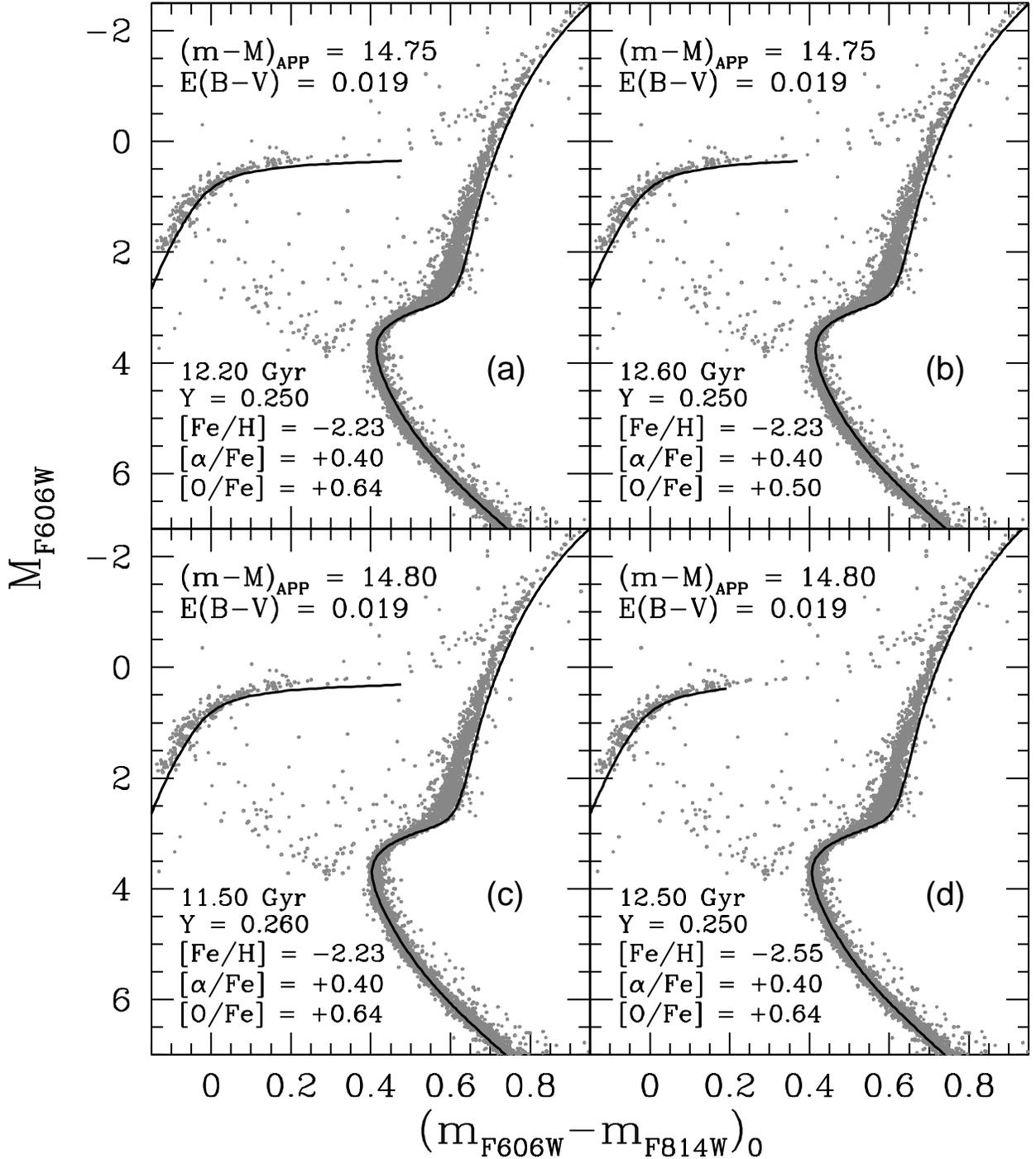}
\caption{The age of M\,92 as inferred from isochrones for the indicated
chemical abundances that produce the best matches to the observed turnoff
luminosity if $E(B-V) = 0.019$ (from SF11) and the apparent distance moduli
are based on fits of ZAHB models to the cluster HB.  The models in panel (a)
assume the abundances that have been derived for HD\,140283, whereas those
in panel (b) adopt a lower [O/Fe] value by 0.14 dex.  In both of these cases,
the isochrones are $\sim 0.01$--0.015 mag too red along the main sequence.
No such problem is found if either a slightly higher helium abundance ($0.26$,
see panel c) or a higher [Fe/H] value ($-2.55$, see panel d) is assumed.  Note
that the reddest ZAHB model in each panel has a mass that is identical to the
turnoff mass of the best-fit isochrone: less massive ZAHB models are bluer.
As discussed in the text, the redward extent of a ZAHB is a strong function of
the assumed iron and oxygen abundances, especially at low [Fe/H] values.}
\label{fig:fig6}
\end{figure}

Suppose the ZAHB-based distance is the correct one and that M\,92 and
HD\,140283 are chemically indistinguishable.  As shown in panel (a) of
Figure~\ref{fig:fig6}, a 12.20 Gyr isochrone for the derived {\it initial}
abundances of HD\,140283 provides quite a good match to the MS and TO of
M\,92, aside from being slightly too red (by only $\approx 0.01$ mag).
To make this age determination, isochrones for different ages were adjusted in
the horizontal direction by whatever amount was needed in order to match the
turnoff color: the one that provided the best fit to the cluster stars from
$\sim 1$ mag below the TO to $\sim 0.6$ mag above it was taken to be the best
estimate of the turnoff age.  A thorough discussion and justification of this
procedure is provided by \citet{vbl13}, who obtained a higher age by about 0.5
Gyr because the models that they employed assumed slightly lower values of
[Fe/H] and [O/Fe] and their ZAHB models yielded a smaller distance modulus by
0.03 mag (which, by itself, accounts for about a 0.3 Gyr age difference).  If
M\,92 has the same [Fe/H] as HD\,140283, but a somewhat lower oxygen
abundance (say, [O/Fe] $= +0.50$ instead of $+0.64$), the same distance modulus
is obtained, but the predicted age increases to $\approx 12.6$ Gyr (see panel
b).  This isochrone also follows along the red edge of the cluster CMD.  To
obtain an age near 13.0 Gyr, the models would need to assume [O/Fe] $\lta 0.3$.

Note that the ZAHB for higher [O/Fe] extends to redder colors.  In these and
all similar plots presented in this paper, the mass of the reddest ZAHB model
is identical to the mass at the tip of the RGB that is predicted by the
isochrone which is fitted to the turnoff photometry, without taking any mass
loss whatsoever into account.  The intrinsic color of the reddest ZAHB star
in M\,92 appears to be $\approx 0.20$.  At this color, the ZAHB models shown
in panels (a) and (b) have masses of 0.730 and 0.748 $\msol$, respectively,
whereas the corresponding RGB tip (and TO) masses are 0.790 and 0.797 $\msol$.
Thus, relatively little mass loss appears to be necessary to explain the 
reddest ZAHB stars in M\,92, though $> 0.2 \msol$ must be lost to explain
the bluest stars (assuming that mass loss is the primary cause of the dispersion
in color along the HB).  The decreased importance of mass loss, compared with
the predictions of older models, is one of the consequences of the reduced rate
of the $^{14}$N$(p,\,\gamma)^{15}$O reaction (\citealt{mfg08}).  As shown by
\citet{pcs10}, this revision causes a ZAHB model for a fixed mass and chemical
abundances to be shifted to significantly higher temperatures --- an effect 
that is especially noticeable at low metal abundances.  

The small offset between the predicted and observed MS and TO colors that are
apparent in panels (a) and (b) can be eliminated if models for either a slightly
higher helium abundance ($Y = 0.260$ instead of 0.250) or a somewhat lower
metallicity ([Fe/H] $= -2.55$ rather than $-2.23$) are overlaid onto the
observed CMD (see panels c and d).  To be sure, $\teff$, color, and other
uncertainties are large enough that this is no more than suggestive.  Still,
the improved agreement leads one to wonder if \citet{vbl13} generally found it
necessary to shift isochrone (but not ZAHB) colors to the blue by typically
$\sim 0.02$ in their large survey of GC ages because they adopted helium
abundances that were too low, or more likely, [Fe/H] values that were too high.
Although $Y = 0.250$ is within current uncertainties of recent determinations
of the primordial abundance ($0.2485 \pm 0.0016$; \citealt{ksd11}), it is
possible that a somewhat larger value is more appropriate for Population II
stars --- perhaps especially those found in GCs, since some processing of the
gas through the AGB stars of an earlier stellar generation (see, e.g.,
Gratton et al.~2012) or via an initial supermassive star (\citealt{dh14}) is
suggested by the chemical abundance variations found in them. 

\begin{figure}
\plotone{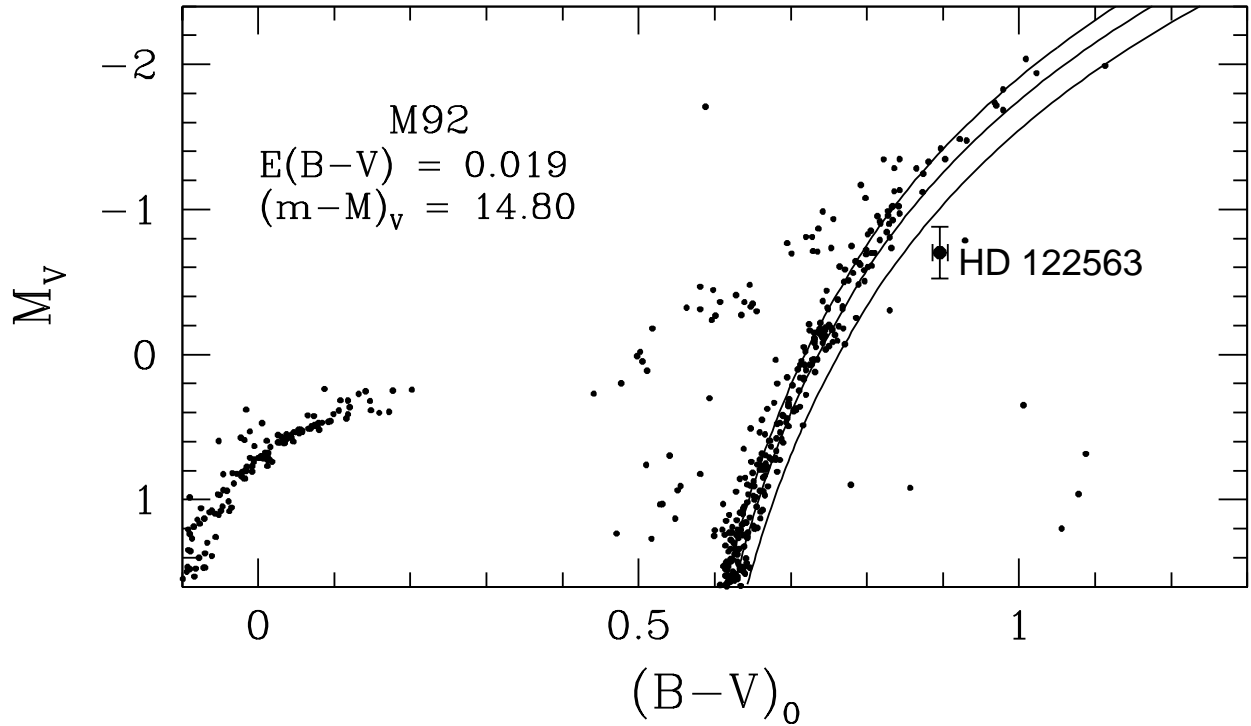}
\caption{Comparison of the HB and upper RGB of M\,92, if the indicated values
of $E(B-V)$ (from SF11) and $(m-M)_V$ (see the previous figure) are assumed,
with the CMD location of the extremely metal-deficient ([Fe/H] $\lta -2.6$)
field giant HD\,122563.  The value of $M_V$, and the associated error bar,
that have been plotted for the latter are based on its {\it Hipparcos} parallax
(\citealt{vl07}); i.e., no Lutz-Kelker correction has been applied (see
footnote 8).   Also shown are 12 Gyr isochrones for [Fe/H] $= -2.8, -2.5$,
and $-2.2$ (the solid curves, in the direction from left to right): each
isochrone also assumes [O/Fe] $= +0.8$, though the location of the RGB does
not depend on the oxygen abundance.}
\label{fig:fig7}
\end{figure}

Insofar as the possibility that M\,92 has [Fe/H] $\lta -2.5$ is concerned,
we note that \citet{rs11} have obtained [Fe/H] $= -2.70 \pm 0.03$ from
a spectroscopic analysis of 19 of its red giants.  Similar investigations by
\citet{pst06} and \citet{sks11} have yielded [Fe/H] $\lta -2.6$ for M\,15,
which has generally been found to have nearly the same iron abundance as M\,92
(see, e.g., \citealt{zw84}, \citealt{cg97}, \citealt{ki03}).  According to
Roederer \& Sneden (also see Sobeck et al.) the 0.3 dex larger [Fe/H] values
that some of the same investigators had obtained previously for both clusters
(e.g., \citealt{spk00}; \citealt{sjk00}) are due to differences in the atomic
data, the adopted model atmosphere grids, and the treatment of Rayleigh
scattering.  Additional support for a reduced metallicity is provided in
Figure~\ref{fig:fig7}, which compares the relative locations of the M\,92 RGB
and the very metal-deficient field giant HD\,122563 ($\pi = 4.22 \pm 0.55$
mas from {\it Hipparcos}; \citealt{vl07}) on the $[(B-V)_0,\,M_V]$-diagram.
HD\,122563 has measured [Fe/H] values that range from $\sim -2.8$ to $\sim
-2.6$ (e.g., \citealt{cds04}, \citealt{rcl10}, \citealt{mgs11}) and it seems to
have normal $\alpha$-element and oxygen abundances for its metallicity
([$\alpha$/Fe] $\approx 0.45$, \citealt{kc12}; [O/Fe] $\gta 0.7$
\citealt{bms03}); consequently, it should lie on the {\it blue} side of the
cluster giant branch if M\,92 has, e.g., [Fe/H] $= -2.35$ (\citealt{cbg09}).
    
If $V-K_S$ colors are plotted instead of $B-V$, the resultant comparison (not
shown) looks qualitatively nearly identical.  (Photometry for HD\,122563 has
been taken from the studies by \citealt{cds04} and \citealt{ctb12}.)  The main
advantage of examining BV observations is that $B-V$ colors are considerably 
more sensitive than $V-K_S$ to metallicity differences at low [Fe/H] values.
Whereas the 12 Gyr isochrones for [Fe/H] $= -2.8, -2.5$, and $-2.2$ that have
been plotted are clearly separated from one another, they are nearly coincident
on the $[(V-K_S)_0,\,M_V]$-diagram.  Why HD\,122563 is offset by such a 
large amount from M\,92 giants, at the same color or at the same luminosity,
is not known, but it may be that its measured parallax, which has a fairly large
uncertainty, is too high.  Consistency with M\,92 (if the latter has [Fe/H]
$\lta -2.6$) would be obtained if the correct parallax of HD\,122563 were $\pi
\approx 3.2$ mas, which is just within the $2\,\sigma$ error bar of the {\it
Hipparcos} value.

Based on the available evidence, we are inclined to adopt [Fe/H] $= -2.5
\pm 0.15$ for M\,92.  The error bar encompasses earlier spectroscopic
determinations near $-2.4$ by, e.g., \citet{ki03} and \citet{cbg09}, and recent
values $\lta -2.6$ that have been either derived for M\,92 (\citealt{rs11}) or
implied by studies of M\,15 (\citealt{pst06}, \citealt{sks11}) and HD\,122563
(as discussed above).  If M\,92 has [Fe/H] $< -2.55$, it would need to have
[O/Fe] $> 0.64$ in order for our ZAHB models to provide a satisfactory match
to the red end of the cluster HB.  As shown in panel (d) of Fig.~\ref{fig:fig6},
the ZAHB model for a mass that is equal to the TO mass of the best-fit
isochrone is predicted to have $(m_{F606W}-m_{F814W})_0 = 0.19$, which is just
barely compatible with the cluster HB (assuming that the few HB stars with
redder colors are evolved stars since they tend to be brighter than a ZAHB;
see the other panels).  The adoption of a lower metallicity would imply an even
bluer ZAHB unless a higher absolute oxygen abundance is assumed: higher [O/H]
values imply younger ages at a given TO luminosity.  Our analysis suggests
that M\,92 is younger than 13 Gyr, and it may even be younger than 12 Gyr if
its stars have a very high oxygen abundance.  Pending further advances in our
understanding, an age near 12.5 Gyr is our current best estimate.  Thus the
subgiants in M\,92 appear to be different from the field subgiant, HD\,140283,
in having both a younger age and (probably) a lower [Fe/H] value.

Before briefly considering the RR Lyrae constraint on cluster distances, the
distance and age of M\,5 will be derived from fits of isochrones and ZAHBs
to its CMD.

\subsection{The Distance and Age of M\,5}
\label{subsec:m5}

\begin{figure}
\plotone{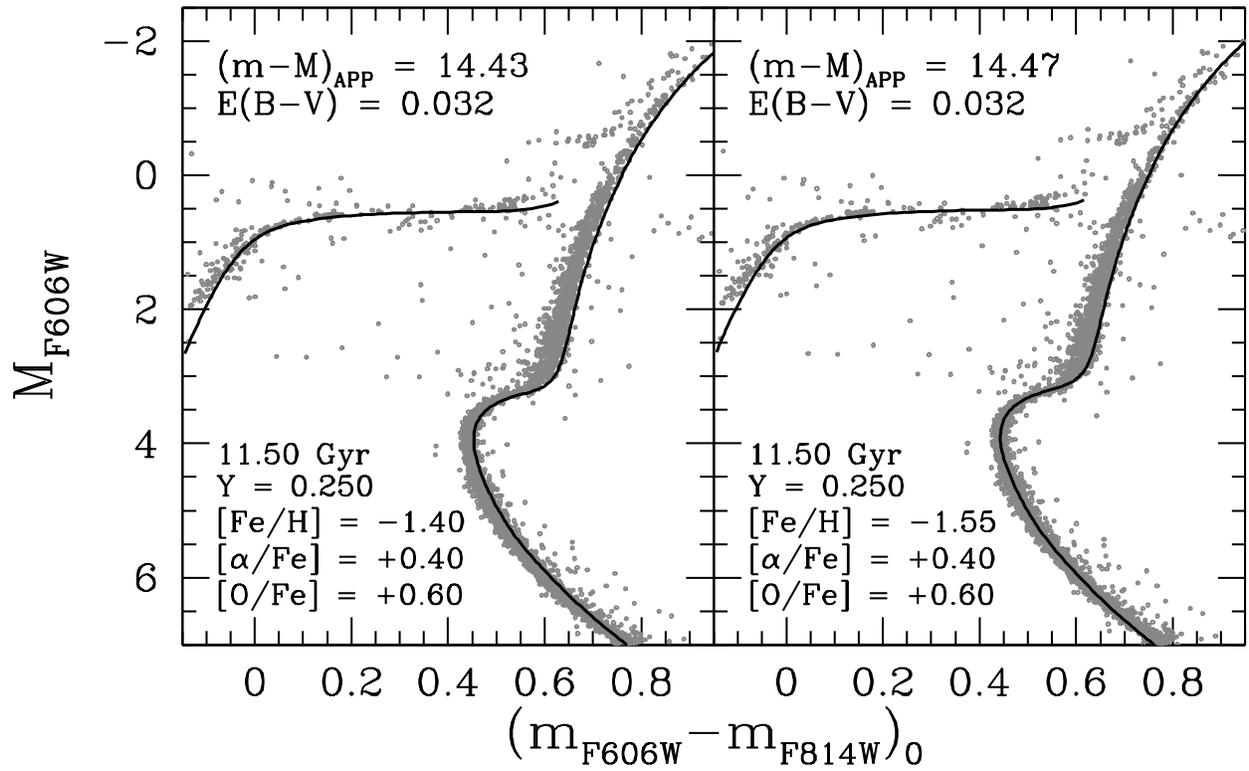}
\caption{Similar to Fig.~6; in this case, two of many possible fits of ZAHB
loci and isochrones for the indicated chemical abundances to the CMD of M\,5
are shown.}
\label{fig:fig8}
\end{figure}

As in the case of M\,92, one could produce many different fits of stellar
models to the CMD of M\,5 given current uncertainties in [Fe/H], [O/Fe], and
$Y$, as well as those associated with the cluster distance and reddening.  We
have chosen to present just two of them.  The left-hand panel of Figure
\ref{fig:fig8} shows that a ZAHB for essentially the derived abundances of
HD\,132475 yields an apparent modulus of 14.43 mag, if $E(B-V) = 0.032$
(from SF11), which implies a TO age of $\approx 11.5$ Gyr.  The ZAHB-based true
distance modulus is thus $\approx 0.10$ mag larger than that inferred from
HD\,132475 (see Fig.~\ref{fig:fig4}) --- a difference that is slightly larger
than the $2\,\sigma$ error bar on the $M_V$ of the field subgiant.  Indeed, this 
explains why the ages that have been determined for M\,5 and HD\,132475
(see Fig.~\ref{fig:fig3}) differ by about 1 Gyr.  [\citet{vbl13} also obtained
an age of 11.5 Gyr for M\,5 using models that assumed the solar abundances
given by \citet{gs98} as the base mixture, with suitably enhanced
$\alpha$-element abundances, and then scaled to [Fe/H] $= -1.33$
(\citealt{cbg09}).  Although they found a smaller modulus by about 0.05 mag,
compared with our estimate, the effects on the age of adopting slightly
larger values of [Fe/H] and [O/H] in their study compensates almost exactly
for this difference.]

The main difficulty with the fit to the observations in the left-hand panel
is that the predicted colors for the MS and TO are too red (though by only
0.01--0.015 mag, which is well within the color uncertainties).  As suggested
in the previous subsection, this discrepancy may be telling us that the adopted
[Fe/H] value is too high.  The right-hand panel shows that the best-fit
isochrone for [Fe/H] $= -1.55$ reproduces the MS and TO photometry of M\,5 very
well without requiring any horizontal offset.  (This is obviously not a
compelling argument for a reduced [Fe/H] value, but neither can one rule out
this possibility given current uncertainties in metal abundance determinations.)
Not unexpectedly, the same age, to within the small fitting uncertainties, is
obtained because both the ZAHB and the TO luminosity at a given age become
brighter as the [Fe/H] value is reduced (though not at identical rates).

\begin{figure}
\plotone{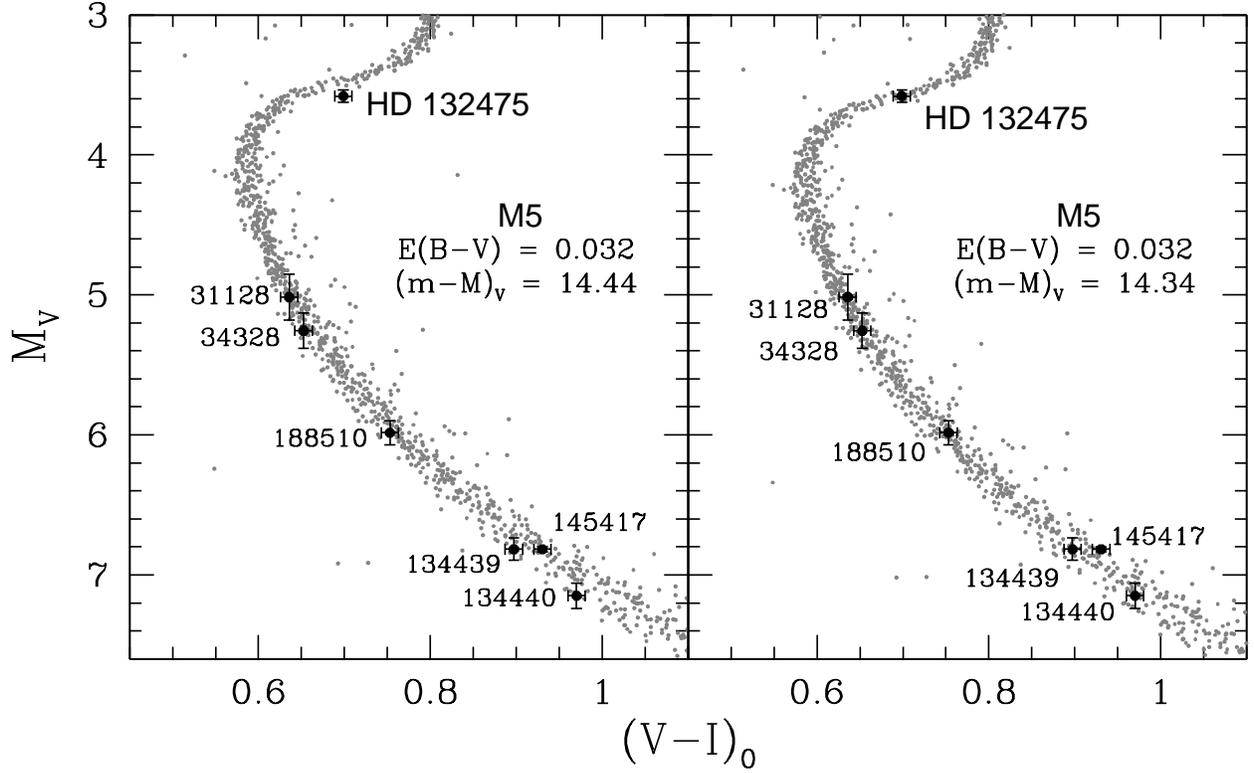}
\caption{Two possible fits of the M$\,$5 CMD to nearby subdwarfs (identified by 
their HD numbers) that have $M_V > 5.0$, $\sigma_\pi/\pi < 0.10$, and [Fe/H]
values that are within $\pm 0.15$ dex of that determined for HD\,132475.  The
vertical error bars are based solely on the parallax uncertainties
(\citealt{vl07}) while $\pm 0.01$ mag has been adopted for the horizontal error
bars.  The only difference between the two panels is the value of $(m-M)_V$
that has been assumed for M\,5 (as indicated).  See the text for the sources of
the photometry, the metallicities, and the adopted reddenings of the subdwarfs.}
\label{fig:fig9}
\end{figure}

However, we can make use of an additional constraint to shed some light on this
issue; namely, subdwarfs in the solar neighborhood with [Fe/H] values similar
to that of M\,5 (and HD\,132475) that also have well determined parallaxes from
{\it Hipparcos}.  In Figure~\ref{fig:fig9}, all of the subdwarfs that we have
been able to identify with $M_V > 5.0$, $\sigma_\pi/\pi < 0.10$
(\citealt{vl07}), and [Fe/H] values within $\pm 0.15$ dex of our determination
for HD\,132475 have been superimposed onto the CMD of M\,5.  For the latter, we
have adopted $E(B-V) = 0.032$ (from SF11) and two different values of the
apparent distance modulus.  The higher value, $(m-M)_V = 14.44$ (left-hand
panel), implies the same true distance modulus as in the left-hand panel of the
previous figure when the differences in $A_V$ and $A_{F606W}$ (see, e.g.,
\citealt{mcc04}) are taken into account.  The lower value, $(m-M)_V = 14.34$
(right-hand panel) is obtained when the CMD of M\,5 is fitted to HD\,132475
(recall Fig.~\ref{fig:fig4}).  The [Fe/H] values, the photometry, and the
adopted reddenings of the subdwarfs have all been taken from the study by
\citet{crm10}, while the M\,5 observations are the same as those employed in
the analysis of the MARCS color--$\teff$ relations by \citet{vcs10}. 

It turns out that the mean [Fe/H] value of the six subdwarfs that satisfied our
criteria is idential to the value that we have derived for HD\,132475 (i.e.,
$-1.51$).  (Due to diffusive effects, the initial [Fe/H] values are expected to
be larger than the observed values by $\sim 0.1$ dex; recall the discussion in
\S~\ref{subsec:initial}.)  Small blueward or redward color offsets ($< 0.01$
mag) were applied to the subdwarf colors, as necessary, to compensate for the
effects of differences between the observed and mean metallicities.  These
corrections were based on the differences in the predicted colors, at the
subdwarf $M_V$ values, of isochrones for the observed range in [Fe/H].
Lutz-Kelker corrections have not been applied to the absolute magnitudes of the
subdwarfs: they are, in any case, quite small (0.03--0.05 mag for the two
brightest stars, $< 0.02$ mag for the others).

Effectively, Fig.~\ref{fig:fig9} is telling us that the apparent distance
modulus of M\,5, as derived from fits of the cluster main-sequence to the sample
of subdwarf calibrators that we have considered is approximately $(m-M)_V = 
14.40$.  (Although not shown, the adoption of a value as low as 14.30 or as high
as 14.50 causes obvious discrepancies between the subdwarfs and the M\,5 CMD.)
While a common age for M\,5 and HD\,132475 is within the uncertainties of the
subdwarf-based distance (see the right-hand panel), distance moduli based on
RR Lyrae stars tend to favor values of $(m-M)_V \sim 14.50$ (e.g.,
\citealt{cdr11}).  For this reason, and because the modulus based on our ZAHB
models is within the uncertainties of the values derived from the subdwarf and
RR Lyrae standard candles (see the next section), we consider $(m-M)_V = 14.45$
to be our ``best estimate" (rounded to the nearest 0.05 mag) of the M\,5
modulus.  In this case, M$\,$5 is predicted to be $\approx 1$ Gyr younger than
HD$\,$132475.

\subsection{The RR Lyrae Standard Candle}
\label{subsec:rrl}

At the present time, the best empirical determination of the slope of the
RR Lyrae $M_V$ versus [Fe/H] relation is $\Delta\,M_V/\Delta\,$[Fe/H] $= 0.214
\pm 0.047$ by \citet{cgb03} from their analysis of $> 100$ variables in the
Large Magellanic Cloud (LMC).  If the true distance modulus of the LMC is
18.50, which is within the 2\% uncertainty of the distance derived
from eight long-period eclipsing binary members by \citet{pgg13}, the mean $V$
magnitude of the RR Lyraes observed by Clementini et al.~($<V_0>$ $= 19.064 \pm
0.064$ at [Fe/H] $= -1.50$) implies that their mean absolute magnitude is
$M_V = 0.564 \pm 0.064$ mag at the reference metallicity.  These results are
plotted in Figure~\ref{fig:fig10} as the {\it open circle} and the {\it dashed
line}.  {\it Dotted lines} have been included in this figure to illustrate the
effect of adopting a steeper or shallower slope by the derived $1\,\sigma$
uncertainty of this quantity.  The findings of \citet{bmf11}, who obtained
$M_V = 0.45 \pm 0.05$ from FGS parallaxes of a small sample of field RR Lyraes,
and of \citet{ksb13}, who derived $M_V = 0.59 \pm 0.10$ using the statistical
parallax technique, are also shown.

According to M.~Catelan (2014, priv.~comm.), the mean apparent magnitudes of the
RR Lyrae variables in M\,5 and M\,92 are 15.062 and 15.081, respectively.
Judging from the results presented in the two previous subsections, M\,5 has
$(m-M)_V = 14.45 \pm 0.07$ and M\,92 has $(m-M)_V = 14.77 \pm 0.07$.
Moreover, [Fe/H] $= -1.40 \pm 0.12$ and $-2.50 \pm 0.15$ appear to be reasonable
estimates of the iron abundances of M\,5 and M\,92, in turn, based on the
available (direct and indirect) evidence.\footnote{It remains to be seen
whether reduced metal abundances will be found for M\,5 when a spectroscopic
analysis is performed that uses the same atomic data, model atmospheres, and
treatment of scattering that led to lower [Fe/H] values by $\sim 0.3$ dex in
the case of M\,15 (\citealt{pst06}, \citealt{sks11}) and M\,92 (\citealt{rs11}).
Until such work is carried out, we are inclined to favor [Fe/H] $= -1.40$ for
M\,5, with an error bar that encompasses the determinations by \citet{zw84},
\citet{ki03}, and \citet{cbg09} and that allows for the possibility of a
somewhat lower value.} Needless to say, it is encouraging to find rather good
consistency of these determinations with the empirical results obtained by
\citet{cgb03} and \citet{ksb13}, to well within their $1\,\sigma$ uncertainties,
and with those reported by \citet{bmf11}, to within $1.5\,\sigma$ (see
Fig.~\ref{fig:fig10}).  Furthermore, when ZAHB-based distance moduli are
adopted, our isochrones, which have been shown to provide good fits to the
local subdwarf calibrators (\citealt{vbl13}), are able to reproduce the
observed main sequences of M\,5 and M\,92 to within $\sim 0.01$ mag in
color (or $\sim 0.05$ mag in magnitude).

As the uncertainties associated with both the RR Lyrae and subdwarf standard
candles are still rather large, it is easily possible that the distance
moduli derived here are too large or too small by several hundredths of a
magnitude.  However, something would have to be seriously wrong with our 
understanding of the HB stars in M\,5 and (especially) M\,92 if the short
distance moduli implied by HD\,132475 and HD\,140283, respectively, are
correct.  Our finding that these two field halo subgiants are significantly
older than well studied GCs of similar chemical compositions seems compelling.

\section{Summary}
\label{sec:sum}

The goal of this investigation since the outset was to obtain improved ages for
HD\,84937, HD\,132475, and HD\,140283, which are the three nearest Population
II subgiants with [Fe/H] $\lta -1.5$ that can be age-dated directly using
trigonometric parallaxes.  To this end, we have used the FGS on the {\it Hubble
Space Telescope} to refine their parallaxes (\S~\ref{sec:pi}), we have carried
out new spectroscopic determinations of their chemical compositions
(\S~\ref{sec:atm}), and we have employed new sets of isochrones
(\S~\ref{sec:models}) to derive our best estimates of their ages
(\S~\ref{sec:ages}).  It turns out that these subgiants have some interesting
implications for our understanding of GCs that have similar metallicities (as
discussed in \S~\ref{sec:gcs}). The main results of this study are as follows:

\begin{enumerate}
\item{Based on 10--11 epochs of FGS observations taken between 2003 August
and 2011 March, we have determined that HD\,84937, HD\,132475, and
HD\,140283 have trigonometric parallaxes $\pi = 12.24 \pm 0.20$, $10.18 \pm
0.21$, and $17.18 \pm 0.26$ mas, respectively.  (By comparison, the {\it
Hipparcos} parallaxes for these stars have larger uncertainties by factors of
$\sim 2.5$--4; see \citealt{vl07}.)  The resultant reddening-corrected absolute
magnitudes are, in turn, $M_V = 3.730 \pm 0.035$, $3.580 \pm 0.045$, and $3.368
\pm 0.033$.}

\item{From analyses of high-resolution, high S/N spectra (with non-LTE
and 3D effects taken into account), we obtained [Fe/H] $= -2.08$ and [O/Fe] $=
+0.44$ for HD\,84937, [Fe/H] $= -1.51$ and [O/Fe] $= +0.61$ for HD\,132475,
and [Fe/H] $= -2.38$ and [O/Fe] $= +0.64$ for HD\,140283.  The measured
abundances of Mg, Si, and Ca indicate that these stars have $\alpha$-element
abundances similar to those typically found in extreme Population II stars
([$\alpha$/Fe] $\approx 0.4$), with the possible exception of HD\,140283, for
which we determined [$\alpha$/Fe] $= 0.26$.  The three subgiants appear to have
close to solar $m$/Fe number abundance ratios of C and N.}

\item{Isochrones were compared with the observed locations of the subgiants on
the $(\log\teff, M_V)$-diagram, resulting in ages of $12.09 \pm 0.14$ Gyr for
HD\,84937, $12.56 \pm 0.46$ Gyr for HD\,132475, and $14.27 \pm 0.38$ Gyr
for HD\,140283 --- where the error bars are based solely on the parallax
uncertainties.  These models were generated for the observed [O/Fe] values,
[$m$/Fe] $= 0.4$ for the other $\alpha$-elements, and the spectroscopic
determinations of [Fe/H] with an adjustment of $+0.10$ dex, in the case of
HD\,132475, and $+0.15$ dex in the case of the other (lower metallicity)
stars, to compensate for the effects of diffusive and extra mixing processes
over their lifetimes; see \citealt{nkr12}, \citealt{gkr13}.)  The age of
HD\,84937 is poorly constrained because it is located just past the turnoff
where isochrones are nearly vertical.  The estimated 60~K $1\,\sigma$ error bar
in its temperature implies $\delta$(age) $= \pm 0.63$ Gyr.  In contrast, the
corresponding age uncertainties for HD\,132475 and HD\,140283 are $\pm
0.26$ and $\pm 0.37$ Gyr, respectively.  However, as discussed by Bond et
al.~(2013, see their Table 1), errors in the derived abundances (particularly
[O/H]), the adopted $V$ magnitudes and reddenings, etc., imply $\sigma$(age)
$\approx 0.8$ Gyr even for the latter two stars.}

\item{As first reported by Bond et al.~(2013), and despite considerable
additional work in the meantime, the age that we have derived for HD\,140283
is slightly greater than the age of the universe as inferred from observations
of the CMB by about 0.5 Gyr, and approximtely 0.7 Gyr older
than the expected maximum age of Population II stars (see \S~\ref{sec:intro}).
However, the associated $1\,\sigma$ error bars overlap.  The most likely
explanations for these difficulties, which would impact our results for the
other subgiants as well, are (i) the absolute oxygen abundance that we have
determined is too low, (ii) the adopted temperature is too low, (iii) the
isochrone $\teff$\ scale is too high, or some combination of these
possibilities.  Alternatively, it remains a remote possibility that HD\,140283
truly is older than 14 Gyr, and that current estimates of the age of the
universe are too low.  In this regard, we note that values of the Hubble
constant which have been determined in some recent investigations do not agree
very well with the values deduced from CMB studies.  For instance, \citet{tr13}
obtained $H_0 = 63.7 \pm 2.3$ km s$^{-1}$ Mpc$^{-1}$ in their investigation of
distant Type Ia supernovae, as compared with the {\it WMAP} value of $H_0 =
69.32 \pm 0.80$ km s$^{-1}$ Mpc$^{-1}$ (\citealt{blw13}). In any case,
HD\,140283 is a very old star that must have formed soon after the Big
Bang.\footnote{As far as we are aware, HD\,140283 does not have any
anomalous properties, such as rapid rotation (e.g., its measured $v\,\sin\,i$
is $2.0 \pm 0.14$ km s$^{-1}$; see \citealt{aln06}) or strong magnetic fields,
that might be partially responsible for an unusually faint $M_V$ (and/or red
color). The simplest interpretation of the observations is that this subgiant is
representative of the very metal poor ([Fe/H] $< -2$) component of the halo of
the Milky Way and that this population is very old.  Nevertheless, in view of
such implications for broader issues, further work on HD\,140283 would clearly
be worthwhile --- including, in particular, an asteroseismic study, which has
the potential to yield interesting independent constraints on its radius and age
(e.g., \citealt{mmt10}).} }

\item{Had we neglected the gravitational settling of helium in the stellar
models that were compared with HD\,140283, the age discrepancies described 
in the previous point would have been $> 1.5$ Gyr.  Thus, models
that neglect this physics appear to be ruled out.  Moreover, diffusive models
must allow for extra mixing just below envelope convection zones in order to
obtain {\it reasonable} consistency between the predicted and observed metal
abundances of TO and lower RGB stars in GCs, and the chemical abundance
variations along the SGB. [The difference between the observed and 
initial [$m$/H] value is not expected to be the same for each metal (as we have
assumed) given that, e.g., some species are more affected by radiative
acclerations and/or turbulent mixing than others (see \citealt{rmr02}).  To
test diffusion physics, it is therefore important to compare the measured
surface abundances of many elements with those predicted for stars in several
evolutionary states.  For instance, \citet{ogk14} have recently found systematic
differences between the observed and predicted metal abundances in stars
belonging to the open cluster M\,67.  Even though such efforts require
exceedingly careful and precise work, they should be given strong support.]}

\item{HD\,140283 appears to be significantly ($\gta 1.5$ Gyr) older than M\,92
(and presumably other GCs of similar metallicities; see \citealt{vbl13}).
If the cluster subgiants that have the same intrinsic color
as HD\,140283 also have the same $M_V$, the apparent distance modulus of
M\,92 would be $(m-M)_V \approx 14.53$.  This would be completely at odds
with expectations based on MS-fits to local subdwarfs or the application of
the RR Lyrae standard candle (or ZAHB-based distance estimates).  HD\,132475
also appears to be older (by $\sim 1$ Gyr) than M\,5, which is believed to
have similar iron and oxygen abundances.  Ages greater than $\approx 12.5$ and
$\approx 11.5$ Gyr that have been derived here for M\,92 and M\,5,
respectively, would be favored if the GCs have significantly lower oxygen
abundances than those found in the two subgiants.  Nonetheless, the Milky Way 
apparently contains field halo stars that are older than those which currently
reside in its GCs.}

\item{Stellar evolutionary models that satisfy the subdwarf and RR Lyrae
constraints to within the (still fairly large) uncertainties associated with
the latter seem to provide the best fits to observed GC CMDs if the low
metallicity end of the cluster [Fe/H] scale is shifted to lower values by
$\sim 0.1$--0.15 dex.  This possibility is supported by recent spectroscopic
work on M\,92 and its near [Fe/H] twin M\,15, as well as the
location of M\,92 giants relative to that of the field giant HD\,122563
([Fe/H] $\lta -2.6$) on the $[(B-V)_0,\,M_V]$-diagram.}

\end{enumerate}
 
The parallaxes reported here are likely to be supplanted later this decade by
precise results from {\it Gaia} (e.g., \citealt{dmc12}).  However, as we have
shown, the parallax error is no longer the dominant contributor to the
uncertainty in the ages of the three subgiants.  In this investigation, the
error bars associated with chemical abundance determinations and the empirical
$\teff$\ scale are comparable to or more important than distance scale
uncertainties.  That will continue to be the case in the coming era when
accurate distances have been measured for vast numbers of stars in the extended
solar neighborhood.

\begin{figure}
\plotone{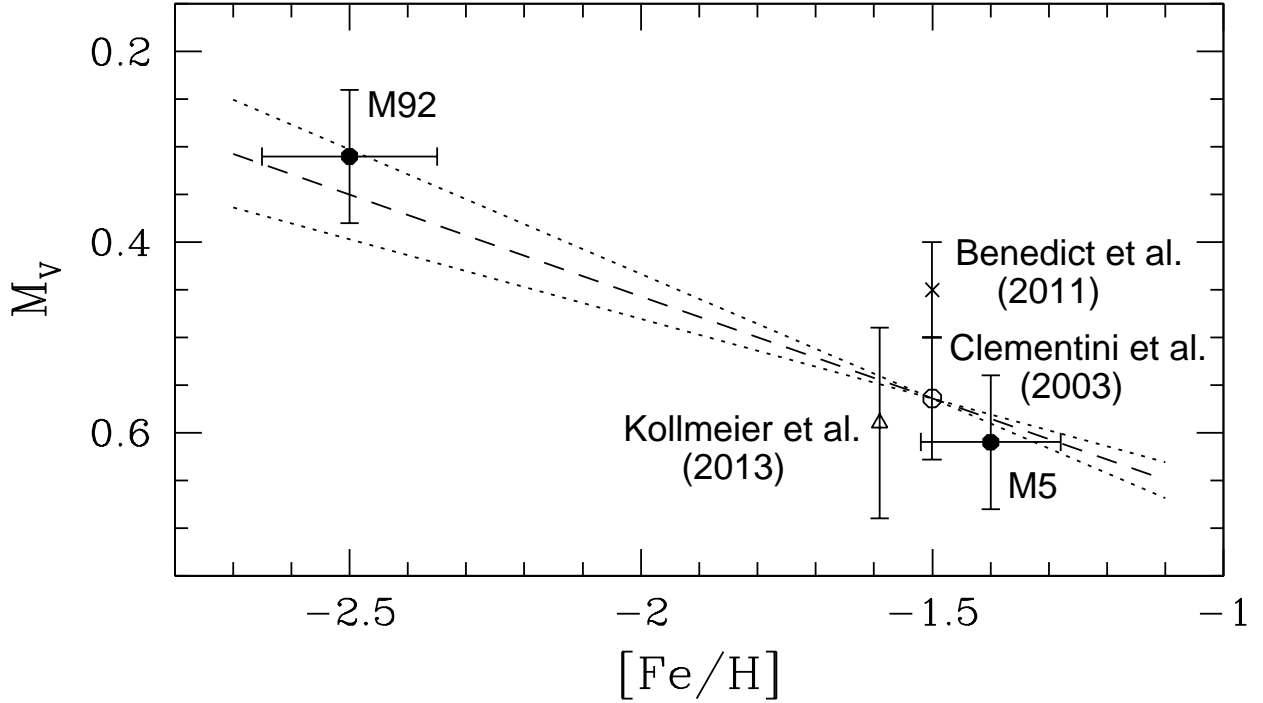}
\caption{Empirical determinations of the mean RR Lyrae luminosity and its
uncertainty, at the reference [Fe/H] $= -1.50$, by Benedict et al.~(2011, {\it
cross}) and Clementini et al.~(2003, {\it open circle}), and at [Fe/H] $= -1.59$
by Kollmeier et al.~(2011, {\it open triangle}).  The results by Clementini et
al., whose determination of the slope of the $M_V$ versus [Fe/H] relation is
shown as a {\it dashed line}, is tied to the true distance modulus of the LMC.
For the latter, a value of 18.50 has been adopted (see \citealt{pgg13}).  The
{\it dotted lines} show the impact of varying the slope by its measured
uncertainty.  The {\it filled circles} represent the locations of M\,92 and
M\,5, assuming our current best estimates of their distance moduli and
metallicities, and the mean apparent magnitudes of their RR Lyraes kindly
provided to us by M.~Catelan (2014, priv.~comm.).}
\label{fig:fig10}
\end{figure}

\acknowledgements
We thank Marcio Catelan for providing the mean apparent magnitudes of the
RR Lyraes in M\,5 and M\,92 that have been used in Fig.~\ref{fig:fig10},
and Volker Bromm for helpful comments concerning the early chemical evolution
of the universe as well as a few pertinent references.  STScI summer students
Ryan Leaman and Mihkel Kama assisted with the reduction of the data for the
reference stars.  DAV is grateful for the support of a Discovery Grant from the
Natural Sciences and Engineering Research Council of Canada.  PEN acknowledges
support from the Stellar Astrophysics Centre, funded by the Danish National
Research Foundation (Grant Agreement No.~DNRF106).  This research used the
facilities of the Canadian Astronomy Data Centre operated by the National
Research Council of Canada with the support of the Canadian Space Agency.
Support for Program number GO-9883 was provided by NASA through a grant from
the Space Telescope Science Institute, which is operated by the Association
of Universities for Research in Astronomy, Incorporated, under NASA contract
NAS5-26555.

\clearpage

\clearpage

\clearpage

\clearpage

\clearpage

\clearpage

\clearpage

\clearpage

\clearpage

\clearpage

\clearpage
\begin{deluxetable}{lllccclcc}
\tabletypesize{\footnotesize}
\tablewidth{0pt}
% \tablewidth{6.5in}
\tablecaption{Astrometric Reference Star Data \label{tab:tab1}}
\tablehead{
\colhead{ID} & \colhead{RA} & \colhead{Dec.} & \colhead{$V$} &
\colhead{$B-V$} & \colhead{$V-I$} & 
\colhead{Sp.~Type} &
\colhead{$\pi_{\rm est}$} & \colhead{$\pi_{\rm FGS}$}\\
\colhead{  } & \colhead{(J2000)} & \colhead{(J2000)} & \colhead{   } &
 \colhead{   } & \colhead{     } & \colhead{       } & \colhead{(mas)} &
\colhead{(mas)} 
}
\startdata
\noalign{\centerline{HD 84937 Field}}\\
\noalign{\vskip -0.3in}\\
HD\,84937-R1  & 09 49 12.43 & +13 41 46.0   & 13.846 & 0.512 & 0.660 & F8 V   &
              $1.47\pm0.29$ & $1.48\pm0.09$ \\
HD\,84937-R2  & 09 49 06.87 & +13 41 56.6   & 13.706 & 0.621 & 0.721 & G0 V   &
              $1.75\pm0.35$ & $1.74\pm0.10$ \\
HD\,84937-R3  & 09 48 55.11 & +13 43 51.9   & 15.557 & 1.251 & 1.487 & K7 V   &
              $3.10\pm0.62$ & $2.70\pm0.16$ \\
HD\,84937-R4  & 09 48 42.28 & +13 46 26.8   & 11.716 & 1.015 & 1.043 & K0 III &
              $1.12\pm0.22$ & $1.22\pm0.07$ \\
HD\,84937-R5  & 09 48 33.21 & +13 46 02.3   & 12.698 & 0.947 & 1.028 & K0 IV  &
              $0.79\pm0.16$ & $0.79\pm0.05$ \\
HD\,84937-R6  & 09 48 34.24 & +13 45 05.1   & 15.051 & 1.162 & 1.185 & K1 III &
              $0.16\pm0.03$ & $0.16\pm0.01$ \\
HD\,84937-R7  & 09 48 25.37 & +13 46 57.5   & 14.820 & 0.809 & 0.883 & K0 V   &
              $1.70\pm0.34$ & $1.29\pm0.06$ \\
\noalign{\vskip -0.1in}\\
\noalign{\centerline{HD 132475 Field}}\\
\noalign{\vskip -0.3in}\\
HD\,132475-R1 & 15 00 05.77 & $-22$ 01 32.4 & 14.999 & 0.752 & 0.932 & G0 V   &
              $1.86\pm0.37$ & $1.86\pm0.08$ \\  
HD\,132475-R2 & 15 00 07.52 & $-22$ 02 11.4 & 15.133 & 0.798 & 0.932 & G6 V   &
              $1.59\pm0.32$ & $1.59\pm0.07$ \\   
HD\,132475-R3 & 14 59 48.56 & $-22$ 01 05.1 & 15.376 & 0.744 & 0.862 & G0 V   &
              $1.09\pm0.21$ & $1.11\pm0.06$ \\   
HD\,132475-R4 & 14 59 49.46 & $-21$ 59 18.2 & 15.623 & 0.663 & 0.820 & G5 V   &
              $0.91\pm0.18$ & $0.91\pm0.04$ \\   
HD\,132475-R5 & 14 59 43.06 & $-21$ 59 20.7 & 14.117 & 0.821 & 0.862 & G5 V   &
              $1.66\pm0.33$ & $1.73\pm0.12$ \\   
HD\,132475-R6 & 14 59 56.34 & $-22$ 00 18.1 & 14.255 & 0.776 & 0.859 & G5 V   &
              $1.68\pm0.34$ & $1.62\pm0.11$ \\   
HD\,132475-R7 & 14 59 30.01 & $-22$ 00 27.3 & 13.959 & 0.629 & 0.731 & G0 V   &
              $1.29\pm0.26$ & $1.28\pm0.06$ \\   
\noalign{\vskip -0.1in}\\
\noalign{\centerline{HD 140283 Field}}\\
\noalign{\vskip -0.3in}\\
HD\,140283-R1 & 15 43 11.66 & $-10$ 55 31.7 & 14.254 & 0.761 & 0.864 & G5 V   &
              $1.59 \pm0.32$ & $1.60\pm 0.19$ \\
HD\,140283-R2 & 15 54 08.74 & $-10$ 54 58.3 & 11.949 & 1.041 & 1.133 & G8 III &
              $1.42 \pm0.28$ & $1.41\pm 0.06$ \\
HD\,140283-R3 & 15 42 59.53 & $-10$ 55 26.3 & 13.174 & 0.812 & 0.887 & G5 IV  &
              $1.12 \pm0.22$ & $1.15\pm 0.08$ \\
HD\,140283-R4 & 15 42 46.83 & $-10$ 57 16.2 & 13.175 & 1.141 & 1.204 & K0 III &
              $0.58 \pm0.12$ & $0.56\pm 0.08$ \\
HD\,140283-R5 & 15 42 46.52 & $-10$ 56 46.9 & 14.673 & 0.825 & 0.904 & G5 V   &
              $1.43 \pm0.29$ & $1.41\pm 0.15$ \\
HD\,140283-VTT & 15 43 08.74 & $-10$ 56 58.1 & 16.623 & 1.114 & 1.299 & K2 V  &
              $1.50 \pm0.30$ & $1.46\pm 0.11$ \\
\enddata								
% \tablenotetext{d}{See discussion of REF-12's reddening in text.}
\end{deluxetable} 

\clearpage
\begin{deluxetable}{lrlllll}
\tabletypesize{\footnotesize}
% \tabletypesize{\tiny}
% \setlength{\tabcolsep}{0.05in}
\tablewidth{0pt}
% \tablewidth{6.5in}
\tablecaption{Astrometric Parameters for the Halo Subgiants \label{tab:tab2}}
\tablehead{
\colhead{Star} &  
\colhead{$\mu_{\alpha, {\rm Hipp}}$} & \colhead{$\mu_{\delta, {\rm Hipp}}$} & 
\colhead{$\pi_{\rm Hipp}$} &
\colhead{$\mu_{\alpha, {\rm FGS}}$} & \colhead{$\mu_{\delta,{\rm FGS}}$} & 
\colhead{$\pi_{\rm FGS}$}  \\
\colhead{    } & 
\colhead{($\rm mas\,yr^{-1}$)}& \colhead{($\rm mas\,yr^{-1}$)}& 
\colhead{(mas)}&
\colhead{($\rm mas\,yr^{-1}$)}& \colhead{($\rm mas\,yr^{-1}$)}& 
\colhead{(mas)} 
}
\startdata
\noalign{\vskip -0.1in}\\
HD\,84937  &  $+373.05\pm0.91$ & $-774.38\pm0.33$ & $13.74\pm0.78$ &
             $+377.85\pm0.26$ & $-772.67\pm0.25$ & $12.24\pm0.20$ \\
HD\,132475 &  $-558.49\pm0.85$ & $-500.37\pm0.68$ & $10.23\pm0.84$ &
             $-559.50\pm0.28$ & $-503.22\pm0.21$ & $10.18\pm0.21$ \\ 
HD\,140283 &  $-1114.93\pm0.62$ & $-304.36\pm0.74$ & $17.16\pm0.68$ &
             $-1111.02\pm0.26$ & $-304.76\pm0.17$ & $17.18\pm0.26$ \\
\enddata
\end{deluxetable} 

\clearpage
\begin{deluxetable}{cccrcccccccccccc}
\tabletypesize{\footnotesize}
\tablewidth{510pt}
\tablecaption{Line Data, Measured Equivalent Widths, and Atmospheric LTE
Abundances Derived From 1D MARCS Models \label{tab:tab3}}
\tablewidth{0pt}
\tablehead{ &  &  &  &  & \multicolumn{2}{c}{HD\,19445\tablenotemark{a}} &
 & \multicolumn{2}{c}{HD\,84937\tablenotemark{b}} &
 & \multicolumn{2}{c}{HD\,132475\tablenotemark{c}} &
 & \multicolumn{2}{c}{HD\,140283\tablenotemark{d}} \\
\colhead{ID} & \colhead{$\lambda$} & \colhead{$\chi_{\rm exc}$} &
\colhead{$\log\,gf$} & & \colhead{$EW$} & \colhead{$A({\rm X})$} &
 & \colhead{$EW$} & \colhead{$A({\rm X})$} & & \colhead{$EW$} & 
 \colhead{$A({\rm X})$} & & \colhead{$EW$} & \colhead{$A({\rm X})$} \\
 &  \colhead{(\AA )}    &  \colhead{(eV)}  & &   &  \colhead{(m\AA }) & & &
 \colhead{(m\AA )}  &  & &  \colhead{(m\AA )}  & &  &  \colhead{(m\AA )}  & }
%\noalign{\smallskip}
%\hline\hline
%\noalign{\smallskip}
%\noalign{\smallskip}
%\hline
%\noalign{\smallskip}
\startdata
%\noalign{\smallskip}
O\,{\sc i}   & 7771.95 & 9.15  &  0.37   &&13.9 & 7.36 &&20.8 & 7.29 &&33.5 &
   7.98 && 9.0 & 7.14 \\
O\,{\sc i}   & 7774.18 & 9.15  &  0.22   && 9.5 & 7.31 &&14.5 & 7.22 &&26.6 &
   7.96 && 6.2 & 7.09 \\
O\,{\sc i}   & 7775.40 & 9.15  &  0.00   && 7.0 & 7.38 && 9.1 & 7.20 &&17.1 &
   7.90 && 3.7 & 7.07 \\
\noalign{\smallskip}
Fe\,{\sc ii} & 5197.58 & 3.23  & $-$2.22 &&     &      &&13.4 & 5.34 &&40.9 &
   5.92 &&10.2 & 5.01 \\
Fe\,{\sc ii} & 5234.63 & 3.22  & $-$2.18 &&     &      &&15.9 & 5.38 &&44.7 &
   5.95 &&12.7 & 5.07 \\
Fe\,{\sc ii} & 5264.81 & 3.33  & $-$3.13 &&     &      && 2.4 & 5.52 &&10.5 &
   6.09 && 1.8 & 5.22 \\
Fe\,{\sc ii} & 5284.11 & 2.89  & $-$3.11 &&     &      && 4.1 & 5.35 &&17.9 &
   5.92 && 3.2 & 5.02 \\
Fe\,{\sc ii} & 5325.56 & 3.22  & $-$3.16 &&     &      && 2.1 & 5.39 && 8.8 &
   5.93 && 1.7 & 5.11 \\
Fe\,{\sc ii} & 5414.08 & 3.22  & $-$3.58 &&     &      &&     &      && 3.6 &
   5.93 &&     &      \\
Fe\,{\sc ii} & 5425.26 & 3.20  & $-$3.22 &&     &      && 2.0 & 5.41 && 8.6 &
   5.95 && 1.4 & 5.06 \\
Fe\,{\sc ii} & 5534.85 & 3.24  & $-$2.75 &&     &      && 4.8 & 5.36 &&19.0 &
   5.92 && 4.3 & 5.13 \\
Fe\,{\sc ii} & 6084.11 & 3.20  & $-$3.79 &&     &      &&     &      && 2.5 &
   5.93 &&     &      \\
Fe\,{\sc ii} & 6149.25 & 3.89  & $-$2.69 && 1.2 & 5.38 && 1.9 & 5.44 && 5.6 &
   5.87 && 1.1 & 5.07 \\
Fe\,{\sc ii} & 6238.39 & 3.89  & $-$2.60 && 1.8 & 5.46 && 1.6 & 5.28 && 7.9 &
   5.94 &&     &      \\
Fe\,{\sc ii} & 6247.56 & 3.89  & $-$2.30 && 3.2 & 5.42 && 3.5 & 5.32 &&14.2 &
   5.93 && 2.1 & 4.96 \\
Fe\,{\sc ii} & 6432.68 & 2.89  & $-$3.57 && 1.6 & 5.44 && 1.4 & 5.29 && 7.6 &
   5.90 && 1.4 & 5.06 \\
Fe\,{\sc ii} & 6456.39 & 3.90  & $-$2.05 && 5.3 & 5.40 && 6.0 & 5.33 &&     &
      && 4.2 & 5.03 \\
\enddata
%\noalign{\smallskip}
%\hline
%\end{tabular}
%
%\begin{list}{}{}
\tablenotetext{a}{Results are based on spectra from ESO program 65.L-0131.} 
\tablenotetext{b}{Programs 73.D-0024, 80.D-0347, and 82.B-0610.} 
\tablenotetext{c}{Program 65.L-0507.} 
\tablenotetext{d}{Programs 65.L-0131 and 80.D-0347.}
\end{deluxetable}

\clearpage
\begin{deluxetable}{llcccccccc}
\tabletypesize{\footnotesize}
\tablewidth{500pt}
\tablecaption{Atmospheric Parameters and Observed Elemental Abundances
 \label{tab:tab4}}
\tablewidth{0pt}
%\noalign{\smallskip}
%\hline\hline
%\noalign{\smallskip}
\tablehead{ \colhead{Star} & \colhead{$T_{\rm eff}$} & \colhead{$\log\,g$} &
 \colhead{$A({\rm O})$} & \colhead{$A({\rm O})$} &  
 \colhead{$A({\rm Fe})$} & \colhead{$A({\rm Fe})$} &
 \colhead{[Fe/H]} & \colhead{[O/Fe]}  & 
 \colhead{[$\alpha$/Fe]\tablenotemark{a}} \\
 & & & \colhead{\tt{(1D,LTE)}} & \colhead{\tt{(3D,NLTE)}} &
 \colhead{\tt{(1D,LTE)}} & \colhead{\tt{(3D,NLTE)}} & & & }
%\noalign{\smallskip}
%\hline
%\noalign{\smallskip}
\startdata
 HD\,19445  & 6136\,K & 4.43 &  7.35 & 7.24 & 5.42 & 5.47 & $-2.03$ & 0.58 &
   0.39 \\
 HD\,84937  & 6408    & 4.05 &  7.24 & 7.05 & 5.37 & 5.42 & $-2.08$ & 0.44 &
   0.38 \\
 HD\,132475 & 5746    & 3.80 &  7.95 & 7.79 & 5.94 & 5.99 & $-1.51$ & 0.61 &
   0.45 \\
 HD\,140283 & 5797    & 3.70 &  7.10 & 6.95 & 5.07 & 5.12 & $-2.38$ & 0.64 &
   0.26 \\
\enddata
%\noalign{\smallskip}
%\hline
%\end{tabular}
%\begin{list}{}{}
\tablenotetext{a}{This refers to the mean abundance of Mg, Si, and Ca.}
%\end{list}
\end{deluxetable}

\clearpage
\begin{deluxetable}{lccccccc}
\tabletypesize{\footnotesize}
\tablewidth{250pt}
\tablecaption{Photometry\tablenotemark{a}, Reddenings\tablenotemark{b}, and
 Absolute Magnitudes \label{tab:tab5}}
\tablewidth{0pt}
\tablehead{\colhead{Star} & & \colhead{$V$} & \colhead{$B-V$} & \colhead{$V-I$} &
 \colhead{$E(B-V)$} & & \colhead{$M_V$} }
\startdata
\noalign{\smallskip}
  HD\,84937  & & 8.306 & 0.396 & 0.547 & 0.005 & & $3.730 \pm 0.035$ \\
  HD\,132475 & & 8.563 & 0.537 & 0.708 & 0.007 & & $3.580 \pm 0.045$ \\
  HD\,140283 & & 7.205 & 0.487 & 0.683 & 0.004 & & $3.368 \pm 0.033$ \\
\enddata
\tablenotetext{a}{From Casagrande et al.~(2010), who give $\sigma(V) = 0.02$,
 $\sigma$(color) $= 0.008$.}
\tablenotetext{b}{From Mel\'endez et al.~(2010), except for HD\,140283 (see
 \S~\ref{subsec:teff}).}
\end{deluxetable}

\clearpage
\begin{deluxetable}{lcccc}
\tabletypesize{\footnotesize}
\tablewidth{250pt}
\tablecaption{Masses and Ages \label{tab:tab6}}
\tablewidth{0pt}
\tablehead{\colhead{Star} & \colhead{${\cal M}/\msol$} & \colhead{Age} &
 \colhead{$\delta$(Age)\tablenotemark{a}} & 
 \colhead{$\delta$(Age)\tablenotemark{b}} \\
  &  & \colhead{(Gyr)} & \colhead{(Gyr)} & \colhead{(Gyr)} }
\startdata
\noalign{\smallskip}
  HD\,84937  & 0.78 & 12.09 & $\pm 0.14$ & $\pm 0.63$ \\
  HD\,132475 & 0.80 & 12.56 & $\pm 0.46$ & $\pm 0.26$ \\
  HD\,140283 & 0.75 & 14.27 & $\pm 0.38$ & $\pm 0.37$ \\
\enddata
\tablenotetext{a}{Derived from the $M_V$ error bar.}
\tablenotetext{b}{Derived from the $\log\,T_{\rm eff}$ error bar.}
\end{deluxetable}


\newpage
 
\begin{thebibliography}{}

\bibitem[Ade et al.(2013)]{aaa13}
Ade, P.~A.~R., Aghanim, N., Armitage-Caplan, C., et al.~2013, arXiv:1303.5076

\bibitem[Allende Prieto, Lambert, \& Asplund(2001)]{ala01}
Allende Prieto, C., Lambert, D.~L., \& Asplund, M.~2001, ApJL, 556, L63

\bibitem[Alves-Brito et al.(2010)]{ama10}
Alves-Brito, A., Mel\'endez, J., Asplund, M., Ram\'irez, I., and Yong,
 D.~2010, A\&A, 513, A35

\bibitem[Asplund(2005)]{asp05}
Asplund, M.~2005, ARA\&A, 43, 481

\bibitem[Asplund et al.(2009)]{ags09}
Asplund, M., Grevesse, N., Sauval, A.~J., \& Scott, P.~2009, ARA\&A, 47,
  481

\bibitem[Asplund et al.(2006)]{aln06}
Asplund, M., Lambert, D.~L., Nissen, P.~E., Primas, F., \& Smith, V.~V.~2006,
 ApJ, 644, 229

\bibitem[Barbuy et al.(2003)]{bms03}
Barbuy, B., Mel\'endez, J., Spite, M., et al.~2003, ApJ, 588, 1072

\bibitem[Barklem(2007)]{bar07}
Barklem, P.~S.~2007, A\&A, 462, 781

\bibitem[Benedict et al.(2011)]{bmf11}
Benedict, G.~F., McArthur, B.~E., Feast, M.~W., et al.~2011, AJ, 142, 187

\bibitem[Benedict et al.(2009)]{bmn09}
Benedict, G.~F., McArthur, B.~E., Napiwotzki, R., et al.~2009, AJ, 138, 1969

\bibitem[Bennett et al.(2013)]{blw13}
Bennett, C.~L., Larson, D., Weiland, J.~L., et al.~2013, ApJS, 208, 20

\bibitem[Bergemann(2008)]{ber08}
Bergemann, M.~2008, Phys.~Scr., 133, 014013

\bibitem[Bessell(2007)]{b07}
Bessell, M.~S.~2007, PASP, 119, 605

\bibitem[Bessell \& Norris(1982)]{bn82}
Bessell, M.~S., \& Norris, J.~1982, ApJL, 263, L29

\bibitem[Bond \& MacConnell(1971)]{bm71}
Bond, H.~E., \& MacConnell, D.~J.~1971, ApJ, 165, 51

\bibitem[Bond et al.(2013)]{bnv13}
Bond, H.~E., Nelan, E.~P., VandenBerg, D.~A., Schaefer, G.~H., \& Harmer,
 D.~2013, ApJL, 765, L12

\bibitem[Carbon et al.(1987)]{cbk87}
Carbon, D.~F., Barbuy, B., Kraft, R., Friel, E.~D., \& Suntzeff, N.~1987,
 PASP, 99, 335

\bibitem[Carretta et al.(2009)]{cbg09}
Carretta, E., Bragaglia, A., Gratton, R.~G., D'Orazi, V., \& Lucatello, 
 S.~2009, A\&A, 508, 695

\bibitem[Carretta et al.(2010)]{cbg10}
Carretta, E., Bragaglia, A., Gratton, R.~G., Recio-Blanco, A., Lucatello, S.,
 D'Orazi, V., \& Cassisi, S.~2010, A\&A, 516, A55

\bibitem[Carretta \& Gratton(1997)]{cg97}
Carretta, E., \& Gratton, R.~G.~1997, A\&AS, 121, 95

\bibitem[Carretta et al.(2000)]{cgc00}
Carretta, E., Gratton, R.~G., Clementini, G., \& Fusi Pecci, F.~2000, ApJ, 533,
 215

\bibitem[Casagrande et al.(2010)]{crm10}
Casagrande, L., Ram\'irez, I., Mel\'endez, J., Bessell, M., \& Asplund, 
  M.~2010, A\&A, 512, A54

\bibitem[Casagrande \& VandenBerg(2014)]{cv14}
Casagrande, L., \& VandenBerg, D.~A.~2014, MNRAS, in press

\bibitem[Cayrel et al.(2004)]{cds04}
Cayrel, R., Depagne, E., Spite, M., et al.~2004, A\&A, 416, 1117

\bibitem[Chang \& Tang(1990)]{ct90}
Chang, T.~N., \& Tang, X.~1990, JQSRT, 43, 407

\bibitem[Clementini et al.(2003)]{cgb03}
Clementini, G., Gratton, R.~G., Bragaglia, A., Carretta, E., Di Fabrizio, L.,
 \& Maio, M.~2003, AJ, 125, 1309

\bibitem[Collet, Asplund, \& Trampedach(2007)]{cat07}
Collet, R., Asplund, M., \& Trampedach, R.~2007, A\&A, 469, 687

\bibitem[Coppola et al.(2011)]{cdr11}
Coppola, G., Dall'Ora, M., Ripepi, V., et al.~2011, MNRAS, 416, 1056

\bibitem[Creevey et al.(2012)]{ctb12}
Creevey, O.~L., Th\'evenin, F., Boyajian, T.~S., et al.~2012, A\&A, 545, A17

\bibitem[Dean, Warren, \& Cousins(1978)]{dwc78}
Dean, J.~F., Warren, P.~R., \& Cousins, A.~W.~J.~1978, MNRAS, 183, 569

\bibitem[Dekker et al.(2000)]{ddk00}
Dekker, H., D'Odorico, S., Kaufer, A., Delabre, B., \& Kotzlowski, H.~2000,
 Proc.~SPIE, 4008, 434

\bibitem[Denissenkov \& Hartwick(2014)]{dh14}
Denissenkov, P., \& Hartwick, F.~D.~A.~2014, MNRASL, 437, L21

\bibitem[Denissenkov \& VandenBerg(2003)]{dv03}
Denissenkov, P., \& VandenBerg, D.~A.~2003, ApJ, 593, 509

\bibitem[di Cecco et al.(2010)]{dbb10}
di Cecco, A., Becucci, R., Bono, G., et al.~2010, PASP, 122, 991

\bibitem[Dopita et al.(2012)]{dmc12}
Dopita, M.~A., Marconi, M., Clementini, G., \& Brocato, E.~2012, ApSS, 341, 1

\bibitem[Drawin(1968)]{dra68}
Drawin, H.-W.~1968, Z.~f\"ur Physik, 211, 404

\bibitem[Dotter et al.(2010)]{dsa10}
Dotter, A., Sarajedini, A., Anderson, J., et al.~2010, ApJ, 708, 698

\bibitem[Fabbian et al.(2009a)]{fab09}
Fabbian, D., Asplund, M., Barklem, P.~S., Carlsson, M., \& Kiselman, D.~2009a,
 A\&A, 500, 1221

\bibitem[Fabbian et al.(2009b)]{fna09}
Fabbian, D., Nissen, P.~E., Asplund, M., Pettini, M., \& Akerman, C.~2009b,
 A\&A, 500, 1143

\bibitem[Formicola et al.(2004)]{fic04}
Formicola, A., Imbriani, G., Costantini, H., et al.~2004, Phys.~Let.~B, 591,
  61

\bibitem[Gehren et al.(2004)]{gls04}
Gehren, T., Liang, Y.~C., Shi, J.~R., Zhang, H.~W., \& Zhao, G.~2004, A\&A,
 413, 1045

\bibitem[Gonz\'alez Hern\'andez \& Bonifacio(2009)]{ghb09}
Gonz\'alez Hern\'andez, J.~I., \& Bonifacio, P.~2009, A\&A, 497, 497

\bibitem[Gratton, Carretta, \& Bragaglia(2012)]{gcb12}
Gratton, R.~G., Carretta, E., \& Bragaglia, A.~2012, A\&AR, 20, 50

\bibitem[Grevesse \& Sauval(1998)]{gs98}
Grevesse, N., \& Sauval, A.~J.~1998, Sp.~Sci.~Rev., 85, 161

\bibitem[Gruyters et al.(2013)]{gkr13}
Gruyters, P., Korn, A.~J., Richard, O., et al.~2013, A\&A, 555, A31

\bibitem[Gustafsson et al.(2008)]{gee08}
Gustafsson, B., Edvardsson, B., Eriksson, K., et al.~2008, A\&A, 486, 951

\bibitem[Hansen et al.(2013)]{hka13}
Hansen, B.~M.~S., Kalirai, J.~S., Anderson, J., et al.~2013, Nature, 500, 51

\bibitem[Hanson(1979)]{han79}
Hanson, R.~B.~1979, MNRAS, 186, 875

\bibitem[Hibbert et al.(1991)]{hbg91}
Hibbert, A., Bi\'emont, E., Godefroid, M., \& Vaeck, N.~1991, J.~Phys.~B,
 24, 3943

\bibitem[Imbriani et al.(2004)]{icf04}
Imbriani, G., Costantini, H., Formicola, A., et al.~2004, A\&A, 420, 625

\bibitem[Israelian et al.(2004)]{ier04}
Israelian, G., Ecuvillon, A., Rebolo, R., et al.~2004, A\&A, 421, 469

\bibitem[Jeffreys, Fitzpatrick, \& McArthur(1988)]{jfa88}
Jefferys, W.~H., Fitzpatrick, M.~J., \& McArthur, B.~E.~1988,
  Celestial Mechanics, 41, 39

\bibitem[Jonsell et al.(2005)]{jeg05}
Jonsell, K., Edvardsson, B., Gustafsson, B., Magain, P., Nissen, P.~E., \&
 Asplund, M.~2005, A\&A, 440, 321
 
\bibitem[Kirby \& Cohen(2012)]{kc12}
Kirby, E.~N., \& Cohen, J.~G.~2012, AJ, 144, 168

\bibitem[Kollmeier et al.(2013)]{ksb13}
Kollmeier, J.~A., Szczygiel, D.~M., Burns, C.~R., et al.~2013, ApJ, 775, 57

\bibitem[Komatsu et al.(2011)]{ksd11}
Komatsu, E., Smith, K.~M., Dunkley, J., et al.~2011, ApJS, 192, 18

\bibitem[Kraft \& Ivans(2003)]{ki03}
Kraft, R.~P., \& Ivans, I.~I.~2003, PASP, 115, 143

\bibitem[Landolt(1992)]{lan92}
Landolt, A.~U.~1992, \aj, 104, 340

%\bibitem[Layden et al.(2005)]{lsv05}
%Layden, A.~C., Sarajedini, A., von Hippel, T., \& Cool, A.~M.~2005, ApJ, 632,
% 266%

\bibitem[Lind, Bergemann, \& Asplund(2012)]{lba12}
Lind, K., Bergemann, M., \& Asplund, M.~2012, MNRAS, 427, 50

\bibitem[Lind et al.(2008)]{lkb08}
Lind, K., Korn, A.~J., Barklem, P.~S., \& Grundahl, F.~2008, A\&A, 490, 777

\bibitem[Lutz \& Kelker(1973)]{lk73}
Lutz, T.~E., \& Kelker, D.~H.~1973, PASP, 85, 573

\bibitem[Magic, Weiss, \& Asplund(2014)]{mwa14}
Magic, Z., Weiss, A., \& Asplund, M.~2014, astro-ph: 1403.1062

\bibitem[Marta et al.(2008)]{mfg08}
Marta, M., Formicola, A., Gy\"urky, Gy., et al.~2008, Phys.~Rev.~C, 78, 2802

\bibitem[Mashonkina et al.(2011)]{mgs11}
Mashonkina, L., Gehren, T., Shi, J.-R., Korn, A.~J., \& Grupp, F.~2011, A\&A,
 528, A87

\bibitem[Mashonkina, Korn, \& Przybilla(2007)]{mkp07}
Mashonkina, L., Korn, A.~J., \& Przybilla, N.~2007, A\&A, 461, 261

\bibitem[Mayor et al.(2003)]{mpq03}
Mayor, M., Pepe, F., Queloz, D., et al.~2003, The Messenger, 114, 20

\bibitem[McCall(2004)]{mcc04}
McCall, M.~L.~2004, AJ, 128, 2144

\bibitem[Mel\'endez \& Barbuy(2009)]{mb09}
Mel\'endez, J., \& Barbuy, B.~2009, A\&A, 497, 611

\bibitem[Mel\'endez et al.(2010)]{mcr10}
Mel\'endez, J., Casagrande, L., Ram\'irez, I., Asplund, M., \& Schuster,
 W.~J.~2010, A\&A, 515, L3

\bibitem[Melis et al.(2013)]{mrm13}
Melis, C., Reid, M.~J., Mioduszewski, A.~J., Stauffer, J.~R., \& Bower,
  G.~C.~2013, IAU Symposium, 289, 60

\bibitem[Metcalfe et al.(2010)]{mmt10}
Metcalfe, T.~S., Monteiro, M.~J.~P.~F.~G., Thompson, M.~J., et al.~2010,
 ApJ, 723, 1583

\bibitem[Nelan \& Bond(2013)]{nb13}
Nelan, E.~P., \& Bond, H.~E.~2013, \apjl, 773, L26 

\bibitem[Nelan \& Makidon(2002)]{nm02}
Nelan, E., \& Makidon, R.~2002, {\it FGS Data Handbook}, Version 4.0 
 (Baltimore: STScI)

\bibitem[Nissen et al.(2007)]{naa07}
Nissen, P.~E., Akerman, C., Asplund, M., et al.~2007, A\&A, 469, 319

\bibitem[Nissen et al.(2004)]{nca04}
Nissen, P.~E., Chen, Y.~Q., Asplund, M., \& Pettini, M.~2004, A\&A, 415, 993

\bibitem[Nissen et al.(2014)]{ncc14}
Nissen, P.~E., Chen, Y.~Q., Carigi, L., Schuster, W.~J., \& Zhao, G.~2014,
 arXiv: 1406.5218

\bibitem[Nissen et al.(2002)]{npa02}
Nissen, P.~E., Primas, F., Asplund, M., \& Lambert, E.~L.~2002, A\&A, 390, 235

\bibitem[Nissen \& Schuster(2010)]{ns10}
Nissen, P.~E., \& Schuster, W.~J.~2010, A\&A, 511, L10

\bibitem[\"Onehag, Gustafsson, \& Korn(2014)]{ogk14}
\"Onehag, A., Gustafsson, B., \& Korn, A.~2014, A\&A, 562, A102

\bibitem[Nordlander et al.(2012)]{nkr12}
Nordlander, T., Korn, A.~J., Richard, O., \& Lind, K.~2012, ApJ, 753, 48

\bibitem[Paxton et al.(2011)]{pbd11}
Paxton, B., Bildsten, L., Dotter, A., Herwig, F., Lesaffre, P., \& Timmes,
 F.~2011, ApJS, 192, 3

\bibitem[Pereira, Asplund, \& Kiselman(2009)]{pak09}
Pereira, T.~M.~D., Asplund, M., \& Kiselman, D.~2009, A\&A, 508, 1403

\bibitem[Perryman(2009)]{per09}
Perryman, M.~2009, {\it Astronomical Applications of Astrometry: Ten Years of
 Exploitation of the Hipparcos Satellite Data} (Cambridge: Cambridge
 Univ.~Press)

\bibitem[Perryman et al.(1997)]{plk97}
Perryman, M.~A.~C., Lindegren, L., Kovalevsky, J., et al.~1997, A\&A, 323, L49

\bibitem[Pietrinferni et al.(2010)]{pcs10}
Pietrinferni, A., Cassisi, S., \& Salaris, M.~2010, A\&A, 522, A76

\bibitem[Pietry\'nski et al.(2013)]{pgg13}
Pietry\'nski, G., Graczyk, D., Gieren, N., et al.~2013, Nature, 495, 76

\bibitem[Piotto et al.(2007)]{pba07}
Piotto, G., Bedin, L.~R., Anderson, J., et al.~2007, ApJL, 661, L53

\bibitem[Pont et al.(1998)]{pmt98}
Pont, F., Mayor, F., Turon, C., \& VandenBerg, D.~A.~1998, A\&A, 329, 87

\bibitem[Preston et al.(2006)]{pst06}
Preston, G.~W., Sneden, C., Thompson, I.~B., Shectman, S.~A., \& Burley,
 G.~S.~2006, AJ, 132, 85

\bibitem[Proffitt \& VandenBerg(1991)]{pv91}
Proffitt, C.~R., \& VandenBerg, D.~A.~1991, ApJS, 77, 473

\bibitem[Ram\'irez, Allende Prieto, \& Lambert(2013)]{ral13}
Ram\'irez, I., Allende Prieto, C., \& Lambert, D.~L.~2013, 764, 78

\bibitem[Ram\'irez et al.(2010)]{rcl10}
Ram\'irez, I., Collet, R., Lambert, D.~L., Allende Prieto, C., \& Asplund,
 M.~2010, ApJ, 725, L223

\bibitem[Ram\'irez, Mel\'endez, \& Chanam\'e(2012)]{rmc12}
Ram\'irez, I., Mel\'endez, J., \& Chanam\'e, J.~2012, ApJ, 757, 164

\bibitem[Richard et al.(2002)]{rmr02}
Richard, O., Michaud, G., Richer, J., Turcotte, S., Turck-Chi\`eze, S., \&
 VandenBerg, D.~A.~2002, ApJ, 568, 979

\bibitem[Ritter et al.(2012)]{rgs12}
Ritter, J.~S., Safranek-Shrader, C., Gnat, O., Milosavljevi\'c, M., \&
 Bromm, V.~2012, ApJ, 761, 56

\bibitem[Roederer \& Sneden(2011)]{rs11}
Roederer, I.~U., \& Sneden, C.~2011, AJ, 142, 22

\bibitem[Roeser, Demleitner, \& Shilbach(2010)]{rds10}
Roeser, S., Demleitner,  M., \& Schilbach, E.~2010, \aj, 139, 2440 

\bibitem[Safranek-Shrader, Milosavljevi\'c, \& Bromm(2014)]{smb14}
Safranek-Shrader, C., Milosavljevi\'c, M., \& Bromm, V.~2014, MNRASL, 440, L76 

\bibitem[Sandquist(2004)]{san04}
Sandquist, E.~L.~2004, \mnras, 347, 101

\bibitem[Sarajedini et al.(2007)]{sbc07}
Sarajedini, A., Bedin, L.~R., Chaboyer, B., et al.~2007, AJ, 133, 1658

\bibitem[Schlafly \& Finkbeiner(2011)]{sf11}
Schlafy, E.~F., \& Finkbeiner, D.~P.~2011, ApJ, 737, 103\ \ \ (SF11)

\bibitem[Shi et al.(2009)]{sgm09}
Shi, J.~R., Gehren, T., Mashonkina, L., \& Zhao, G.~2009, A\&A, 503, 533

\bibitem[Smith et al.(2013)]{smh13}
Smith, G.~H., Modi, P.~N., \& Hamren, K.~2013, PASP, 125, 1287

\bibitem[Smith \& Raggett(1981)]{sr81}
Smith, G., \& Raggett, D.~St.~J.~1981, J.~Phys.~B., 14, 4015

\bibitem[Sneden et al.(2000)]{sjk00}
Sneden, C., Johnson, J., Kraft, R.~P., Smith, G.~H., Cowan, J.~J., \& Bolte,
 M.~2000, ApJ, 536, L85

\bibitem[Sneden, Pilachowshi, \& Kraft(2000)]{spk00}
Sneden, C., Pilachowski, C.~A., \& Kraft, R.~P.~2000, AJ, 120, 1351

\bibitem[Sobeck et al.(2011)]{sks11}
Sobeck, J.~A., Kraft, R.~P., Sneden, C., et al.~2011, AJ, 141, 175

\bibitem[Soderblom et al.(2005)]{snb05} 
Soderblom, D.~R.,  Nelan, E., Benedict, G.~F., et al.~2005, \aj, 129, 1616

\bibitem[Spergel et al.(2003)]{svp03}
Spergel, D.~N., Verde, L., Peiris, H.~V., et al.~2003, ApJS, 148, 175.

\bibitem[Spite \& Spite(1982)]{ss82}
Spite, F., \& Spite, M.~1982, A\&A, 115, 357

\bibitem[Takeda \& Takada-Hidai(2013)]{tt13}
Takeda, Y., \& Takada-Hidai, M.~2013, PASJ, 65, 65

\bibitem[Tammann \& Reindl(2013)]{tr13}
Tammann, G.~A., \& Reindl, B.~2013, A\&A, 549, A136

\bibitem[Trampedach \& Stein(2011)]{ts11}
Trampedach, R., \& Stein, R.~F.~2011, ApJ, 731, 78

\bibitem[Valcarce, Catelan, \& Sweigart(2012)]{vcs12}
Valcarce, A.~A.~R., Catelan, M., \& Sweigart, A.~V.~2012, A\&A, 547, A5

\bibitem[VandenBerg(2000)]{van00}
VandenBerg, D.~A.~2000, ApJS, 129, 315

\bibitem[VandenBerg et al.(2012)]{vbd12}
VandenBerg, D.~A., Bergbusch, P.~A., Dotter, A., et al.~2012, ApJ, 755, 15

\bibitem[VandenBerg et al.(2014)]{vbf14}
VandenBerg, D.~A., Bergbusch, P.~A., Ferguson, J.~W., \& Edvardsson, B.~2014,
 ApJ, submitted

\bibitem[VandenBerg, Casagrande, \& Stetson(2010)]{vcs10}
VandenBerg, D.~A., Casagrande, L., \& Stetson, P.~B.~2010, AJ, 140, 1020

\bibitem[VandenBerg et al.(2013)]{vbl13}
VandenBerg, D.~A., Brogaard, K., Leaman, R., \& Casagrande, L.~2013, ApJ,
  775, 134

\bibitem[VandenBerg et al.(2002)]{vrm02}
VandenBerg, D.~A., Richard, O., Michaud, G., \& Richer, J.~2002, ApJ, 571, 487

\bibitem[VandenBerg et al.(2000)]{vsr00}
VandenBerg, D.~A., Swenson, F.~J., Rogers, F.~J., Iglesias, C.~A., \&
  Alexander, D.~R.~2000, ApJ, 532, 430

\bibitem[van Leeuwen(2007)]{vl07}
van Leeuwen, F.~2007, A\&A, 474, 653

\bibitem[Zacharias et al.(2013)]{zfg13}
Zacharias, N., Finch, C.~T., Girard, T.~M., et al.~2013, \aj, 145, 44 

\bibitem[Zhao \& Gehren(2000)]{zg00}
Zhao, G., \& Gehren, T.~2000, A\&A, 362, 1077

\bibitem[Zinn \& West(1984)]{zw84}
Zinn, R., \& West, M.~J.~1984, ApJS, 55, 45

\end{thebibliography}
\end{document}